# Failure Diagnosis in Microservice Systems: A Comprehensive Survey and Analysis

SHENGLIN ZHANG, Nankai University, Tianjin, China and Haihe Laboratory of Information Technology Application Innovation (HL-IT), Tianjin, China

SIBO XIA, WENZHAO FAN, BINPENG SHI, XIAO XIONG, and ZHENYU ZHONG, Nankai University, Tianjin, China

MINGHUA MA, Microsoft, Redmond, USA

YONGQIAN SUN*, Nankai University, Tianjin, China and Tianjin Key Laboratory of Software Experience and Human Computer Interaction (TKL-SEHCI), Tianjin, China

DAN PEI, Tsinghua University, Beijing, China

Widely adopted for their scalability and flexibility, modern microservice systems present unique failure diagnosis challenges due to their independent deployment and dynamic interactions. This complexity can lead to cascading failures that negatively impact operational efficiency and user experience. Recognizing the critical role of fault diagnosis in improving the stability and reliability of microservice systems, researchers have conducted extensive studies and achieved a number of significant results. This survey provides an exhaustive review of 98 scientific papers from 2003 to the present, including a thorough examination and elucidation of the fundamental concepts, system architecture, and problem statement. It also includes a qualitative analysis of the dimensions, providing an in-depth discussion of current best practices and future directions, aiming to further its development and application. In addition, this survey compiles publicly available datasets, toolkits, and evaluation metrics to facilitate the selection and validation of techniques for practitioners.

## 1 INTRODUCTION

In the era of the Internet, a multitude of Web applications are emerging, accompanied by a rapid proliferation of diverse device terminals. However, due to the evolving business requirements and the expansion of business scale, the task of maintaining and updating monolithic architecture applications has become increasingly arduous. Microservice systems have emerged as the latest paradigm in constructing modern applications [1]. As a pivotal industry in the digital economy, they are playing a significant role in infusing new vitality into innovation and development across various domains. In the event of failures and deviations from their intended behavior, microservices can experience performance degradation or even system crashes, thereby adversely impacting user experience and resulting in substantial financial losses for enterprises. According to a report [1], an outage lasting 24 hours of mission-critical services from AWS us-east-1 can lead

---

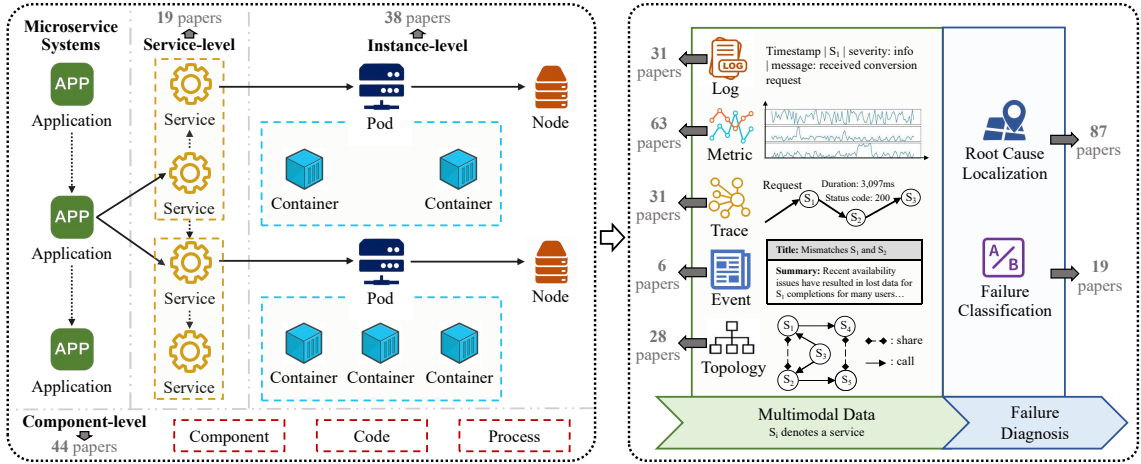

Fig. 1. An overall architecture for failure diagnosis.

to a direct revenue loss of $3.4 billion, while an outage lasting 48 hours can exacerbate the financial impact to reaching $7.8 billion. In 2023 alone, notable service providers such as Microsoft [2], Google [3], and Alibaba Cloud [4] encountered noteworthy failures and downtime incidents. Therefore, to ensure the performance and reliability of microservice systems and minimize losses for businesses, it is imperative to monitor and effectively diagnose performance issues in these systems.

However, the highly heterogeneous topology and diverse interactions of microservice systems pose challenges for operations personnel in formulating general rules or writing manual scripts to promptly localize and classify failures. When addressing the task of failure diagnosis in microservice systems, numerous researchers have explored the optimization and transformation of existing techniques, as well as delving into novel intelligent failure diagnosis solutions. Consequently, there has been a burgeoning demand for artificial intelligence for IT operations (AIOps) [2–4]. AIOps leverages artificial intelligence techniques, including machine learning and deep learning, to autonomously analyze failures, construct models, pinpoint the root cause, and classify the type of failure based on extensive multimodal data (*i.e.*, logs, metrics, traces, events, and topology). In contrast to traditional rule-based and script-based solutions, intelligent failure diagnosis provides enhanced adaptability and generalizability. It represents a novel approach to transcending bottlenecks in traditional solutions and has emerged as one of the mainstream research directions in the field of failure diagnosis in microservice systems.

The failure diagnosis task in microservice systems is inherently intricate and critically significant, leading to a plethora of proposed solutions. However, these solutions are widely dispersed across diverse literature and predominantly focus on either root cause localization or failure classification. Additionally, they often employ varying terminologies for identical concepts or utilize identical terms for distinct concepts. Consequently, this situation poses challenges for practitioners aiming to comprehend failure diagnosis and propose further advancements, as well as complicates the retrieval and aggregation of relevant studies. Due to the interdisciplinary and pervasive nature of this field, coupled with the aforementioned issues, prior studies and surveys have been incomplete, as summarized in Table 1, with the following gaps: (1) **Inadequate analysis and distinction of failure diagnosis objectives**: Previous works [5–9] have lacked a comprehensive analysis of failure diagnosis objectives, particularly in differentiating between the tasks of root cause localization and failure classification. Furthermore, explicit problem statement has been insufficient, resulting in misconceptions and confusion among practitioners. (2) **Limited exploration of multimodal data**: Prior studies [5–12] have primarily focused on techniques that handle single-modal or homogeneous data, leading to a lack of thorough exploration and analysis of techniques that utilize multimodal data, including logs, metrics, traces, and topologies. (3) **Insufficient qualitative analyses from a practical application perspective**: Previous studies [8, 10, 11, 13] have not adequately conducted qualitative analyses from a practical application standpoint, hindering practitioners' understanding of real-world requirements, challenges, and the availability of valuable experiential guidance. (4) **Lack of systematic research and consolidation of publicly available resources** [6–10, 12, 13]: There is a significant absence of systematic research and consolidation concerning publicly available datasets, toolkits, and evaluation metrics, impeding researchers' access to relevant data, suitable tools, and the establishment of standardized criteria for practical implementation and performance assessment.

Table 1. State-of-art survey conducted in fields covered by this work, based on their *objective* (*e.g.*, RCL for root cause localization, and FC for failure classification), the *data* they use (*e.g.*, Single for single-modal data, and Multi for Multimodal data), the *analysis* (*e.g.*, Gra. for granularity, Exp. for explainability, Cha. for characteristic, Por. for portability, and Acc. for accuracy), the publicly available *resource* (*e.g.*, DS for datasets, TK for toolkits, and EM for evaluation metrics).

| Reference | Year | Objective | | Data | | Analysis | | | | | | Resource | | |
|---|---|---|---|---|---|---|---|---|---|---|---|---|---|---|
| | | RCL | FC | Single | Multi | Gra. | Exp. | Cha. | Por. | Acc. | Costs | DS | TK | EM |
| Oliner et al. [10] | 2012 | ✓ | ✓ | ✓ | | ✓ | | | | | ✓ | | | |
| Wong et al. [5] | 2016 | ✓ | | ✓ | | ✓ | | | ✓ | ✓ | | | ✓ | ✓ |
| Gao et al. [6, 7] | 2015 | ✓ | | ✓ | | ✓ | | | | | | | | |
| Sole et al. [8] | 2017 | ✓ | | ✓ | | ✓ | | ✓ | ✓ | ✓ | | | | |
| He et al. [11] | 2021 | ✓ | ✓ | ✓ | | ✓ | | ✓ | ✓ | | | ✓ | ✓ | |
| Li et al. [13] | 2022 | ✓ | ✓ | ✓ | ✓ | ✓ | | | | | | | | |
| Notaro et al. [12] | 2021 | ✓ | ✓ | ✓ | | ✓ | ✓ | ✓ | | ✓ | ✓ | | | ✓ |
| Soldani et al. [9] | 2022 | ✓ | | ✓ | | ✓ | ✓ | ✓ | | ✓ | ✓ | | | |
| **Our work** | - | ✓ | ✓ | ✓ | ✓ | ✓ | ✓ | ✓ | ✓ | ✓ | ✓ | ✓ | ✓ | ✓ |

Therefore, it is imperative to undertake a comprehensive survey and analysis that effectively compiles, categorizes, and summarizes prior contributions to bridge these gaps and offer a more comprehensive understanding of the field.

Leveraging practical insights gleaned from real-world production environments and the aforementioned pertinent studies, as illustrated in Figure 1, we present a primary analysis of 98 papers spanning the past two decades, culminating in the proposition of a comprehensive system architecture for failure diagnosis in microservice systems. Our primary objective is to elucidate the stages of failure diagnosis in microservice systems, furnishing comprehensive introductions, taxonomies, and synopses of prevalent failure diagnosis solutions. Furthermore, we conduct a qualitative analysis of the present progress and prospective future directions and trends, considering aspects such as granularity, explainability, characteristics, portability, accuracy, and costs. To achieve this objective, our contributions can be summarized as follows:

- We undertake a systematic survey of 98 primary papers that specifically focus on failure diagnosis in microservice systems. Subsequently, we meticulously address inconsistencies and ambiguities in terminology and concepts identified across different studies. Building upon this foundation, we propose a comprehensive architecture for failure diagnosis and state the problem, encompassing root cause localization and failure classification.

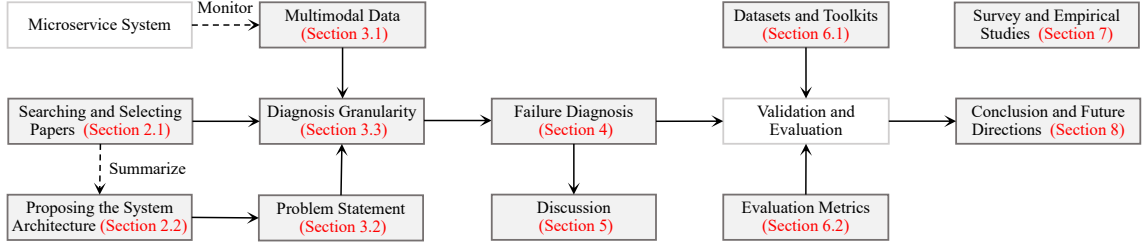

Fig. 2. Overview of topics covered in this article.

- We introduce a comprehensive taxonomy that thoroughly considers the utilized data, properties, core methods, and diagnostic targets of the techniques. Our findings are succinctly summarized in four tables, followed by a qualitative discussion and analysis of the pertinent studies on failure diagnosis. This analysis effectively elucidates key perspectives and requirements, shedding light on the pain points that necessitate attention and improvement. To the best of our knowledge, we are among the pioneering efforts in focusing on failure diagnosis through multimodal data.

- We offer publicly accessible datasets and toolkits that encompass diverse data modalities. Furthermore, we introduce open-source platforms that facilitate practitioners in conducting related research. Additionally, we comprehensively summarize the evaluation metrics for the surveyed techniques. It is noteworthy that none of the existing studies have offered such extensive evaluation support. These contributions hold significant value as they not only aid practitioners in solution selection and validation but also enable quantitative comparisons of failure diagnosis techniques.

We introduce a system architecture and state the failure diagnosis problem to offer a thorough grasp for readers. We conduct an in-depth qualitative analysis of prior research and real-world demands, highlighting best current practices and future directions. Additionally, we provide datasets, toolkits, and metrics for practical use and assessment. We believe that our survey has the potential to catalyze further research endeavors, thereby propelling the field beyond its current boundaries.

The organization of the remainder of this article is illustrated in Figure 2. Section 2 delineates the survey's methodology, providing a concise summary of the architecture. Section 3 introduces the terminology and states the problem of failure diagnosis. Section 4 presents a structured overview of the existing failure diagnosis techniques in microservice systems. Section 6 offers publicly available datasets, toolkits, and evaluation metrics to facilitate failure diagnosis studies. Section 7 elucidates previous related review studies. Section 8 concludes the survey by summarizing the discussion and outlining future directions.

## 2 METHODOLOGY

This survey has been oriented and organized with the intention to answer the following research questions:

**RQ1**: What is the granularity of failure diagnosis in modern microservice systems? How to state the problem of failure diagnosis?

**RQ2**: What are the most important characteristics and methods that define each failure diagnosis technique? Based on this, how to taxonomize each technique and qualitatively analyze each class?

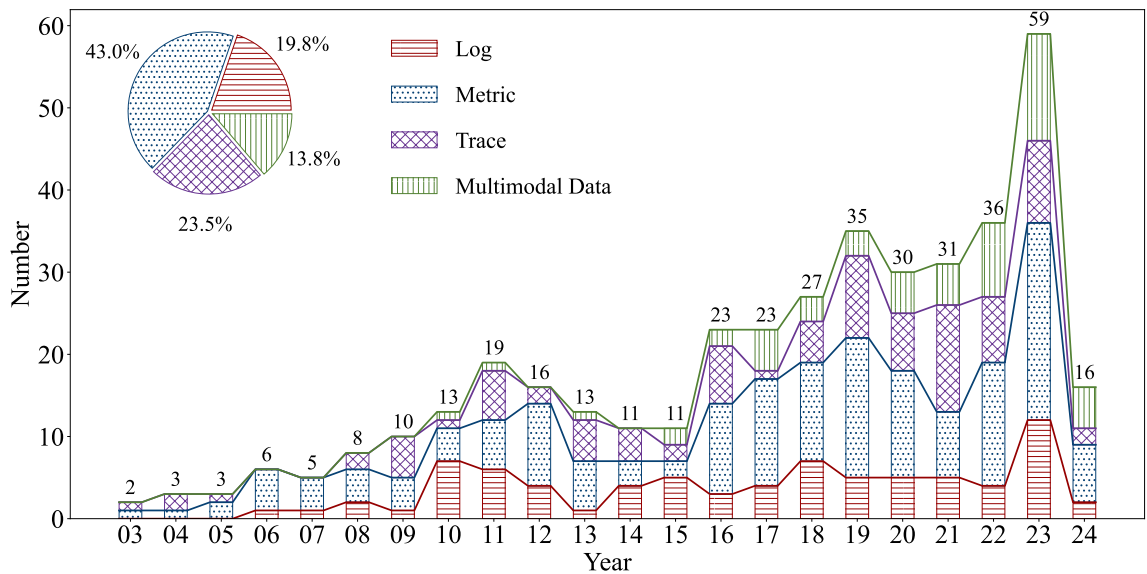

Fig. 3. Publication distribution on each research topic and the associated evolution trend from 2003 to the present.

**RQ3**: Are there publicly available datasets and toolkits for failure diagnosis in the AIOps domain of microservice systems? What are their evaluation metrics?

### 2.1 Study Identification and Selection

To systematically collect papers for conducting this survey, we follow the methodology [14] that provides answers to the proposed research questions.

**Search Strategy.** We first identify the top peer-reviewed and influential conferences and journals in the domains of software engineering, artificial intelligence, security, and data mining. They include 12 conferences (*e.g.*, ASE, ICSE, ISSTA, FSE/ESEC, AAAI, ICML, IJCAI, CCS, NDSS, SIGKDD, ICDE, and VLDB) and 5 journals (*e.g.*, TOSEM, TSE, TDSC, TIFS, and TKDE). We then manually retrieve 73 papers relevant to our objectives published in the past decade. The concept of microservice systems was introduced in 2014 and gradually gained popularity.

In addition, we select 5 scientific and well-known digital repositories, including IEEE Xplore, ACM Digital Library, ScienceDirect, Web of Science, and Scopus. These repositories cover a large number of papers in the domain of computer science, providing convenient search capabilities and rich open information. To search the relevant papers from the above repositories, we mainly use the set of search keywords (see our online appendix [5]) from the relevant papers identified in the conference and journal search. After conducting the keyword search, we collect 2,548 relevant papers spanning from 2003 to the present.

**Selection Criteria.** Post-acquisition of papers, we conduct a relevance assessment based on our pre-defined inclusion and exclusion criteria (see our online appendix [5]). By eliminating papers that are neither in English (exclusion criteria 1) nor innovative in their research (exclusion criteria 2), the total number of papers is reduced to 2,270. Then, we examine the titles, authors, and abstracts of these papers (exclusion criteria 3-4). In total, we collect 400 papers in the domain of

---

failure diagnosis. Figure 3 shows the histogram of annual papers. We can find that the number of papers has steadily increased during this period. The focus on achieving accurate and efficient failure diagnosis in microservice systems has gradually emerged as a key issue of common concern in both academia and industry.

**Quality Assessment.** To highlight representative studies in this domain, we formulate six quality assessment criteria (see our online appendix [5]) to evaluate the effectiveness and significance of the collected papers. We adopt a scoring system ranging from 0 to 3 (poor, fair, good, excellent) for each criterion. After conducting the quality assessment, we obtain 91 papers with total scores exceeding 14 (80% of the maximum possible score).

**Forward and Backward Snowballing.** To ensure comprehensive coverage and avoid missing any relevant studies, we analyze not only the references within these papers but also the publications that have cited these papers. In addition, we repeat the iterative process, which includes applying selection criteria and conducting quality assessment. Finally, we obtain a final set of 98 papers for survey and analysis.

## 2.2 System Architecture

Drawing upon practical experience in real-world production environments and leveraging relevant research on failure diagnosis techniques, we put forth a comprehensive system architecture for diagnosing failures in microservice systems.

Modern microservice systems are commonly constructed on network devices, including switches, routers, firewalls, intrusion detection and prevention systems, and virtual private networks, to establish stable network connections, facilitate data transmission, and ensure robust system operation and data security. As depicted on the left-hand side of Figure 1, to cater to the needs of a substantial user base and process vast quantities of data and requests, contemporary large-scale microservice systems frequently employ tens of thousands of routers and switches to establish connections among hundreds of thousands or even millions of nodes [15]. Each node is equipped with one or more containers, within which the workloads are executed. These containers are grouped as pods, constituting the smallest deployable unit to facilitate creation, scheduling, and management. Given the frequent generation and termination of pods, services provide a level of abstraction and serve as the endpoint for these pods [16]. Industrial microservice applications encompass dozens to thousands of microservices, enabling efficient deployment and orchestration [17]. Owing to the intricate and dynamic nature of the operating environment in microservice systems, when one or more instances of the system experience failure, the failure can propagate gradually to different components or even the entire system, resulting in performance degradation or even service interruption. This poses a substantial risk to the system's stability, necessitating more sophisticated failure diagnosis techniques.

Based on the existing studies on monitoring techniques and tools for microservice systems, we perform an extensive survey and analysis. Failure diagnosis is a crucial task that assists operators in accurately identifying system failures and facilitating swift recovery, ultimately ensuring the reliability and stability of microservice systems. As illustrated on the right-hand side of Figure 1, this architecture primarily outlines and delineates the processes and task scope involved in failure diagnosis. To enable real-time monitoring of the operational status of microservice systems, operators typically engage in continuous collection of five types of observable data: logs, metrics, traces, events, and topology. Among these, logs and metrics represent the primary focus of traditional service monitoring, while traces serve as a dedicated monitoring component that addresses the interaction challenges among microservices, playing a crucial role in microservice system monitoring. They serve as the primary source of information for failure diagnosis, forming the three foundational pillars of observability in contemporary microservice systems [18]. Furthermore, events and topology offer supplementary information, and it is generally insufficient to rely solely on either of them for conducting subsequent failure diagnosis tasks. When a failure occurs, the process of mining correlation relationships and extracting failure features from factors that influence failures is performed. Subsequently, failure diagnosis techniques and models are employed to pinpoint the root cause or classify the failure type as the outcome, facilitating timely resolution and ensuring the smooth operation of the microservice system.

## 3 TERMINOLOGIES

To comprehensively comprehend the constraints of tools and workflows, as well as aid operators in pinpointing the root cause or categorizing the failure type, we extensively survey contemporary literature in modern microservice systems that document and analyze failures occurring in production environments. Initially, this section present the concept of multimodal data for failure diagnosis in microservice systems (Section 3.1). Leveraging the insights gained from the investigated failure diagnosis techniques, this section ultimately provide a comprehensive and detailed statement of the failure diagnosis problem in microservice systems (Section 3.2). Subsequently, we provide a summary of the foundational aspects of failure diagnosis problems, expound upon the level of root cause localization, and meticulously outline the significant failures (Section 3.3).

### 3.1 Multimodal Data

Modern microservice systems exhibit enhanced scalability and accelerated development, fostering innovation and expediting the delivery of novel features, all the while bolstering development agility, operability, scalability, reliability, and the simplicity of monitoring. For instance, when users engage in online chat, they encounter diverse functions, including friend search, interface display, message sending, and message receiving, all of which are facilitated by a dedicated service-oriented application. Owing to the efficient deployment and flexible orchestration of microservices, the interactions among them become intricately complex and dynamically adaptable, thereby posing substantial challenges. To attain precise and efficient failure diagnosis, operational personnel engage in the ongoing collection and monitoring of three discernible data patterns: logs, metrics, and traces. These three data patterns serve as the paramount information sources, comprising the fundamental pillars of observability in microservice systems [18].

**Log.** Logs capture comprehensive events occurring during the runtime of microservice systems, encompassing system information, user behavior information, and business information about network connections. The detailed information within logs exhibits a semi-structured nature and typically comprises fixed template components as well as variable parameter parts. The former encompasses fixed fields that describe system events, such as timestamp, node, service, container, and level. Conversely, the latter records intricate details about system events. Operators generate logs by employing commands like *printf, logging.debug*, and *logging.error* [19]. The abundant semantic information conveyed by logs offers an internal depiction of the system, thereby facilitating diverse tasks related to system management and diagnostics.

**Metric.** Numerous service providers engage in continuous monitoring and recording of metrics for the entire system, aiming to detect anomalous behavior and ensure the elevated quality and dependability of microservices. Metrics enable the measurement of the operational state of services, containers, applications, and other entities. They are stored in the form of a chronologically ordered sequence based on their occurrence, aggregated within predetermined time intervals (*e.g.*, 30 seconds or 1 minute), thereby constituting a data stream. Depending on the entities reflected or indicated by the metrics, they can be categorized into user-perceived metrics (*e.g.*, availability and average response time) and system-level metrics (*e.g.*, CPU utilization and memory utilization). In the event of system failures, metrics swiftly and accurately reflect alterations in system performance and the patterns of diverse failures, thereby assisting operators in diagnosing these failures.

**Trace.** Upon receiving a user request, a microservices system initiates a sequence of invocations among microservices to jointly satisfy the business requirements. Within the OpenTracing standard [6], these invocation processes are denoted as spans, commonly distinguished by *SpanId* and *ParentId*. These identifiers aid in pinpointing the current request's position within the complete business invocation hierarchy and ascertaining its upstream and downstream service nodes. Each span symbolizes a named and timed segment of uninterrupted execution, sharing a distinctive identifier known as *TraceID*. These spans collectively constitute a directed acyclic graph [20, 21]. Programmers commonly integrate the invocation linking interface into the code of each microservice, enabling the recording of invocation relationships among microservices, along with the incorporation of diverse annotations. These annotations typically cover invocation time, request status, latency, and specific information pertinent to various business aspects [22]. These pieces of information depict the execution state of the user request. Unlike logs and metrics, traces can portray the intricate relationships between microservices nodes through the lens of a service dependency graph, thereby yielding more interpretable and granular diagnostic results.

**Event and Topology.** To gain a more comprehensive understanding of the system's operational status and improve the accuracy of failure diagnosis, many works use events [23–28] and topology [1, 16, 19, 24, 29–44] to assist in failure diagnosis. Events refer to records or notifications of significant, meaningful, and impactful situations or conditions within the system. They are usually associated with abnormal behavior, state changes, or important operational operations of the system, which may be triggered by the system, applications, devices, or users, and may include failures, errors, warnings, performance changes, security events, etc. Events are usually represented in structured text form, including basic information such as the event's unique identifier, event type, timestamp, event source, event description, and other related modality information. Topology describes the relationships and connections between various components in the system or network, which can be used to represent system architecture, network structure, application dependencies, and more. Topology is usually represented in the form of a graph, where nodes represent components in the system or network, and edges represent connections between components. Topology can be generated and updated through automatic discovery, monitoring, and analysis to maintain consistency with the system. By analyzing topology, the system can achieve fault troubleshooting, performance optimization, capacity planning, and other functions, helping operators better manage and maintain the system. When using logs, metrics, and traces, integrating events and topology information can help capture system features and states, and achieve more accurate failure diagnosis.

### 3.2 Problem Statement

Consider a large-scale microservice system with services and instances of each service, where monitoring of all system services and instances is achieved by collecting logs, metrics, traces, events, and topology associated with each service and instance. The number of instances for each service can be different. When a failure occurs, operators need to localize the root cause service or instance based on the multimodal data information mentioned above, or further localize the root cause component of the service or instance, determine the failure type, and take timely measures for failure mitigation and repair.

Assuming a failure occurs at a certain timestamp, the first task of failure diagnosis is a ranking problem, where the root cause service or instance ranks higher than other system services or instances, or further localizing the root cause component ranks higher than other components. This task is commonly referred to as root cause localization, which the goal is to estimate the probability of each system service, instance or component being the culprit. The second task

---

is a classification problem, which involves identifying and determining the occurrence of a specific failure type from a predefined set of failure types. This task is typically referred to as failure classification.

### 3.3 Granularity of Failure Diagnosis

We systematically analyzes the collected papers for failure diagnosis in microservice systems. The ultimate target of root cause localization is to provide the most probable root cause when a system failure occurs. Figure 1 shows the possible targets for root cause localization in microservice systems. We conclude that the localization level in the collected papers can be divided into service-level [17, 22, 30, 34, 36, 45–58], instance-level [19–21, 24, 28, 29, 31–33, 35, 37–41, 50, 53, 59–77] and component-level [1, 22, 23, 25–28, 34, 38, 40–44, 50, 53, 54, 65, 73, 78–96]. Modern microservice systems split monolithic applications into multiple independently deployable and runnable services based on specific granularity standards around the business domain. Each node deploys one or more containers. They are bundled together as pods. Each service can be supported with multiple instances, and services can communicate with each other through the network for asynchronous calls. Service-level and instance-level localization can reflect which service or which instance within a service is the root cause of the failure from the perspective of failure diagnosis in microservice systems. Component-level localization provides more granular diagnostic results compared to service-level and instance-level, specifically by not only identifying the system service or instance where the root cause lies but also pinpointing the root cause component within it. This mainly depends on whether the service or instance is treated as a whole.

For failure classification, our survey includes major failures in large-scale microservice systems (*e.g.*, cloud computing platform [39], production microservice system [56], database services [97, 98], OpenStack cloud platform [59], and J2EE PetStore demonstration application[99]), experiences with publicly available microservice systems (*e.g.*, TrainTicket [100]), related studies and industrial situation. Based on the failure manifestation and type description, we collected and organized the following failure types that may occur in the microservice system. Table 2 summarizes these failure types and proposes corresponding mitigation strategies.

(1) Hardware Failures
   - Resource Scheduler Failures [16, 17, 20, 39, 52, 57, 74, 101, 102]. System resource limitations, incorrect scheduling strategies, or improper priority settings can lead to the scheduler's inability to allocate resources correctly. Additionally, due to factors such as heavy system loads, improper resource configurations, or uneven resource allocations, running system components such as hosts, containers, or virtual machines may experience insufficient CPU, memory, disk, or network bandwidth allocations. Moreover, system component failures can also occur as a result of internal damage.
   - Intensive Workload [60, 80, 81, 87, 98, 101, 103]. When the system receives a large number of requests simultaneously or when complex tasks are present, the hardware resources of the database may not be able to meet the I/O or CPU demands in the system, leading to I/O or CPU saturation.
   - Resource Exhaustion [21, 52, 55–57, 75, 80, 84, 104, 105]. CPU consumption and disk errors can be caused by abnormal programs, resulting in busy waiting or deadlock due to competing actions, leading to an infinite loop in data writing. Memory leaks occur when allocated memory blocks are not released after use. Accumulation of unreleased memory can lead to memory shortages and system failures.

(2) Software Errors

Table 2. Typical failure types and mitigation measures in microservice systems.

| Type | Details | Examples | Mitigation Measures |
|---|---|---|---|
| Hardware Failure | Resource Scheduler Failures | System misconfigurations<br>System scheduling exception<br>System resource underprovisioning<br>System component damage | Reconfigure the resource scheduler and scale up the system resource. |
| | Intensive Workload | I/O saturation<br>CPU saturation<br>I/O and CPU saturation | Horizontal scaling adjusts resources and utilizes load-balancing techniques. |
| | Resource Exhaustion | CPU consumption<br>Disk error<br>Memory leak | Implement resource deallocation and capacity planning, along with setting up request throttling and circuit-breaking mechanisms. |
| Software Error | System Bottleneck | Lock contention<br>Process crash<br>Handle leak | Adopt the reasonable resource allocation strategy and resource request sequence. |
| | Poorly Written Query | Incorrect index or join type<br>Execute redundant subqueries<br>Incorrect prepared statements | Avoid redundant data access and optimize query statements. |
| | Poor Physical Design | Incorrect index design<br>Inadequate disk partitioning<br>Incorrect data type design | Optimize and improve the physical design of the database. |
| | Code Bugs | Logic bugs<br>Incorrect data exchange | Use software testing techniques to localize and fix bugs, or perform version rollbacks. |
| | External Operations | System update, migration or upgrade | Flush, backup, and restore the system. |
| Network Problem | Network Exception | Network device breakdown<br>Incorrect network configuration | Check the configuration of the network protocols and the state of the network device. |
| | Transmission Stress | Network congestion<br>Network transmission delay<br>Network transmission abortion | Check the network transmission configuration and reconnect the network transmission. |

- System Bottleneck [17, 19, 25, 53, 87, 97, 98, 106]. When lock contention occurs, it can result in wasted system resources and decreased efficiency. If not promptly resolving lock contention, it can lead to abnormal termination of the application or system crashes. Additionally, handle leaks are also a major cause of performance degradation or crashes in microservice systems.

- Poorly Written Query [58, 97, 102, 105]. Repeated execution of the same subquery, failure to use a *WHERE* clause for filtering, incorrect use of indexes or appropriate join types, and failure to use parameterized queries or prepared statements correctly can result in decreased query performance or incorrect results, while also increasing the risk of SQL injection attacks.

- Poor Physical Design [58, 97]. Improper index design, disk partitioning, and data type design are among the software errors that can lead to system failures. For example, creating too many indexes on a frequently updated database can significantly increase the overhead of write operations. Using data types that are too large or too small can result in wasted storage space or data truncation.

- Code Bugs [74, 98, 102]. Bugs are prevalent in practice. For example, errors in logic or handling in application code can cause it to not run as expected. This includes mistakes in conditional statements, loops, algorithms, and so on. In modern microservice systems, the formats and values of data exchanged between the sender and receiver are often different. This category also includes incorrect data exchange and access permission denied exceptions caused by distributing only partial updates to certificates or credentials.

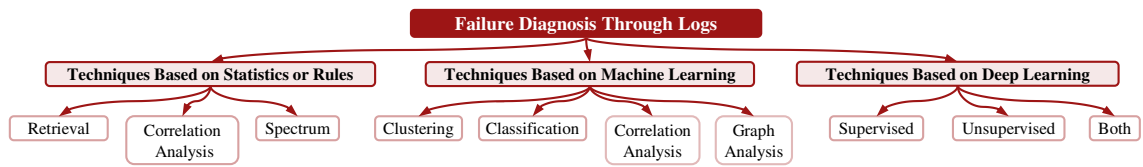

Fig. 4. Categories of failure diagnosis techniques through logs.

- External Operations [59, 84, 98, 106]. In the development of microservice systems, updating or upgrading is a common operation. For example, during database migration or upgrade, writing or executing incorrect scripts can lead to issues such as data corruption, interrupted connections, and incompatible engines.

(3) Network Problems

- Network Exception [16, 39, 76, 87, 88, 98, 103, 104]. Network devices (*e.g.*, routers, switches, firewalls) can experience failures. Additionally, incorrect network configurations such as IP address conflicts, subnet mask misconfigurations, and routing configuration errors can lead to network communication failures or abnormal data transmission.
- Transmission Stress [21, 59, 60, 75, 80, 97, 101, 105, 107]. When a system operates in a high-load environment with simultaneous large-scale data transmission and high-concurrency request processing, issues such as network congestion, network transmission delay, and network transmission abortion can occur due to packet processing delays on gateways, incorrect routing policies, or an inadequate network topology.

## 4 FAILURE DIAGNOSIS

We survey and distinguish failure diagnosis techniques into four classes based on the data they use: logs (Section 4.1), metrics (Section 4.2), traces (Section 4.3), and multimodal data (Section 4.4). Subsequently, we summarize the findings in four summary tables (Table 3, 4, 5, 6) to offer a comprehensive overview of the existing research and practical advancements.

### 4.1 Failure Diagnosis Through Logs

Logs record event information during the runtime of a microservice system, including system data, user behavior, and network-related business information. These logs support various management and diagnostic tasks. However, most failure diagnosis techniques need structured inputs [59, 60, 78, 79, 81–83, 85, 86, 88, 104, 107, 108]. This involves transforming raw logs into parameters and log templates. Parameters include fixed fields like timestamp, node, service, container, and level, while log templates capture detailed system events.

Early designs [109] introduced tools like POD-Discovery and POD-Viz for process model extraction and real-time visualization, aiding operators in understanding process context. However, these manual methods are time-consuming and impractical for massive multi-source logs in microservice systems. As shown in Figure 4, more automated log analysis methods have since emerged [59, 60, 78–89, 104, 107, 108], categorized into techniques based on statistics/rules, machine learning, and deep learning.

*4.1.1 Techniques Based on Statistics or Rules.* Early log analysis techniques used traditional methods based on statistics or rules, which were simple yet effective. These methods are categorized into retrieval [78, 79, 81], correlation analysis [80], and spectrum [82].

**Retrieval.** LOGAN [78] creates a reference model for each request type, representing normal log patterns. For failure diagnosis, current logs are compared with this model to identify root causes. LogDC [79] uses a naive Bayesian network to label deployment logs and compares them with normal logs for failure diagnosis. Unlike LOGAN and LogDC, GLOBECOM'18 [81] actively collects action logs in faulty environments for diagnosis. All these techniques [78, 79, 81] build databases associating logs with failures and diagnose issues by retrieving from these databases.

**Correlation Analysis.** LADRA [80] extracts features from logs, including execution, memory, and CPU-related aspects. It defines seven feature-related factors (*e.g.*, abnormal ratio, memory change rate, loading delay) and uses linear regression to weigh these factors, calculating the probability of root causes based on feature correlation.

**Spectrum.** Spectrum-based failure localization (SBFL) techniques, used in software debugging, identify potential code errors based on test case coverage and results. SBLD [82] applies SBFL to logs, using differences in event occurrences to pinpoint issues. It employs hierarchical agglomerative clustering (HAC) to rank and cluster events, aiding in selecting relevant events for troubleshooting.

*4.1.2 Techniques Based on Machine Learning.* Compared to Section 4.1.1, techniques based on machine learning replace statistical or rule-based methods with more advanced machine learning methods, offering stronger modeling capabilities. Firstly, clustering [83, 84, 104] and classification techniques [85] categorize historical logs into different groups, then perform failure diagnosis through retrieval or matching. Secondly, to identify relationships between features and failures, correlation analysis techniques [86, 87] employ machine learning algorithms like regression trees, dependency networks, Pearson correlation, and time-group heuristics. Lastly, some methods focus on using logs for graph construction [88, 89].

**Clustering.** LogCluster [104] improves upon retrieval-based techniques [78, 79, 81] (Section 4.1.1) by parsing log messages into weighted events using IDF and contrast-based methods. It then applies HAC to group similar sequences and select representatives. During production, LogCluster compares failure sequences to these representatives using cosine similarity, suggesting mitigation actions based on historical data. However, it requires manual labeling of representative sequences, which is time-consuming. To address this, Log3C [83] combines sequence clustering with user-perceived metrics to automatically identify log sequences associated with system performance degradation. It employs cascading clustering to group sequences at various time intervals and uses t-statistics to analyze the relationship between cluster sizes and metrics. This approach accelerates the clustering process for massive sequences through sampling, clustering, pattern extraction, and matching. While Log3C incorporates both logs and metrics, its core still revolves around log clustering, with metrics primarily determining whether a cluster reflects a failure rather than pinpointing specific failure locations.

Onion [83] introduced the concept of diagnostic targets as incident-indicating logs, characterized by consistency, impact, and bilateral difference. These characteristics are incorporated into a progressive clustering process for vectorized logs, resulting in a clustering tree where each node represents a combination of log vectors. To maintain cluster consistency, log vectors are divided into subgroups based on similarity at each tree depth. To mitigate the time-consuming nature of splitting to leaf nodes, Onion implements a downward-closure-based pruning strategy using a popularity ratio to evaluate log vector impact scope, with a threshold to halt node splitting. This ensures each node or cluster in the tree indicates widespread anomalies across servers. Finally, Onion defines a contrast score to rank root causes, considering differences between normal and incident-indicating logs.

**Classification.** LogFaultFlagger [85] aims to maximize failure detection while minimizing root cause log lines to streamline investigations. After parsing and one-hot encoding templates, it uses set-difference with passed logs to isolate failure-specific logs and assigns importance based on IDF. LogFaultFlagger then employs exclusive K nearest neighbors (KNN) to predict if a log line will cause a failure. In essence, if any of a log vector's K neighbors has triggered a bug report, it's deemed likely to cause a failure. This approach aligns with the goal of comprehensive failure identification.

**Correlation Analysis.** DISTALYZER [86] extracts event and state features from logs, describing the system's runtime state qualitatively and quantitatively. It uses t-tests to identify the most discriminative features between normal and abnormal groups. However, significantly different features aren't necessarily root causes, as divergences often increase along interrelated feature chains. DISTALYZER employs dependency networks to establish a graph between features and user-perceived metrics, introducing attention focusing to find a small set of highly divergent features strongly dependent on these metrics. While most existing work only identifies errors leading to failures, FDiagV3 [87] provides more comprehensive failure-related information. It first identifies associated jobs, nodes, correlated events, and event sequences by extracting relevant data from logs to form a matrix. Principal component analysis (PCA) and independent component analysis (ICA) extract outliers, while the Pearson correlation algorithm identifies related log events. FDiagV3 also introduces a time-group heuristics method to construct failure-related event sequences, supplementing event localization results.

**Graph Analysis.** Several techniques utilize logs to construct graphs for analysis. The Peter and Clark (PC) algorithm [110], a popular causal analysis method based on probabilistic graphical models, determines causal relationships by examining conditional independence. It begins with a complete undirected graph and iteratively removes edges that fail to meet conditional independence criteria, resulting in a directed acyclic causality graph. ICWS'17 [88] employs a two-level graph construction approach to model execution states between and within services during normal periods. It first builds a service topology graph using log frequency and the PC algorithm, then creates time-weighted control flow graphs (TCFGs) to describe internal microservice states. TCFGs use log templates as nodes, connecting them with edges based on frequent sequential appearance and short time intervals. Edge weights represent transition times between templates. By comparing these graphs during normal periods, ICWS'17 can diagnose sequence, redundancy, and latency anomalies.

ICWS'17 [88] constructs graphs at both service and event levels. HALO [89] further refines this approach to the attribute field level, utilizing telemetry monitoring data (structured log components resembling attribute fields) for failure diagnosis. Considering the cloud's hierarchical nature, HALO builds an attribute hierarchy graph through pairwise relationship identification, skeleton extraction, and skeleton-based clustering. It then applies a self-adaptive top-down search on failure-aware random walk paths to identify failure-related attribute value combinations. Reverse truncation ensures appropriate granularity in the localization process.

*4.1.3 Techniques Based on Deep Learning.* Deep learning techniques for log analysis can be categorized as supervised [59, 108] or unsupervised [60, 107, 108] based on their supervision methods.

**Supervised.** Cloud'19 [59] focuses on logs from single system tasks for diagnostic analysis. It aggregates tracing information and error messages with the same request ID, using the first log entry's event as the task representative. For log vectorization, Cloud'19 trains a word2vec model on task-specific logs. It then aggregates log vectors by request ID, using the centroid to represent each request. Finally, Cloud'19 trains a supervised classifier for each task to map request vectors to diagnosis reports, including failure types, server nodes, and additional information.

**Unsupervised.** Log IDs represent abstract concepts or specific instances, enabling system workflow reconstruction from logs for detailed failure diagnosis. SwissLog [60] builds an ID relation graph and uses a heuristic algorithm for instance-level localization, scanning nodes to eliminate unlikely root causes. In contrast, LogKG [107] employs knowledge graphs to analyze multiple log fields, including templates, components, and request IDs. Through entity extraction and graph construction, LogKG obtains encodings for each log template, then uses TF-IDF weighting for failure-oriented log representations. For downstream tasks, LogKG applies the OPTICS clustering algorithm to label root causes for each cluster.

Table 3. Summary of surveyed failure diagnosis techniques through logs, based on their *category*, the publication *year*, the *data* (*e.g.*, L for logs, and M for metrics) they use, the *principle* (*e.g.*, R for techniques based on rule, ML for techniques based on machine learning, and DL for techniques based on deep learning) and *supervision* (*e.g.*, Sup for supervised techniques, and Unsup for unsupervised techniques) type, the core *method*, and the diagnostic *target* (*e.g.*, I for instance-level localization, C for component-level localization, and FT for failure types).

| Category | Technique | Year | Data | Principle | Supervision | Core Method | Target |
|---|---|---|---|---|---|---|---|
| Statistics or Rules | LOGAN [78] | 2016 | L | R | - | Log Correlation + Log Parsing and Clustering + Log Alignment | C |
| | LogDC [79] | 2017 | L | R | - | Natural Language Toolkit + Naive Bayesian Network + Template mining + K-means Clustering + Logistic Regression + Likelihood Estimation | C |
| | GLOBECOM'18 [81] | 2018 | L | R | - | Log Template Extraction + Sample-based and Probability Model + Greedy Entropy Minimization + Reinforce Learning | C |
| | LADRA [80] | 2017 | L | R | - | 3-sigma + Weighted Factor + Classical Liner Regression | C |
| | SBLD [82] | 2020 | L | R | - | Log Parsing + SBFL + HAC | C |
| Machine Learning | LogCluster [104] | 2016 | L | ML | Sup | Log Parsing [111] + IDF + HAC + Sigmoid Function + Cosine Similarity | FT |
| | Log3C [83] | 2018 | L, M | ML | Unsup | Log Parsing [111] + IDF + Multivariate Linear Regression Model + Cascading Clustering + HAC + T-statistic | C |
| | Onion [84] | 2021 | L | ML | Unsup | Term-Frequency + Term-Importance + Progressive Clustering + Contrast Score | C |
| | LogFaultFlagger [85] | 2019 | L | ML | Sup | Static Vocabulary + Weighted-IDF + Cosine Similarity + Exclusive KNN | C |
| | DISTALYZER [86] | 2012 | L, M | ML | Unsup | T-tests + Dependency Networks + Regression Trees | C |
| | FDiagV3 [87] | 2015 | L | ML | Unsup | PCA + ICA + Pearson Correlation + Time-group Heuristics | C |
| | ICWS'17 [88] | 2017 | L | ML | Unsup | PC [110] + Probabilistic Model | C |
| | HALO [89] | 2021 | L | ML | Unsup | Pairwise Conditional Entropy + Uncertainty Reduction + Hierarchy Intensity + Skeleton-based Clustering + Failure-aware Random Walk + Self-adaptive Top-down Search | C |
| Deep Learning | Cloud'19 [59] | 2019 | L | ML, DL | Sup | Word2vec + KNN + Naive Bayes + Neural Networks + Random Forest | I, FT |
| | SwissLog [60] | 2022 | L | DL | Unsup | Log Parsing + BERT + Linear Transformation + Attention-based Bi-LSTM + Heuristic Algorithm | I |
| | LogKG [107] | 2023 | L | DL | Unsup | Rule Extraction + Common Event Expression + Knowledge Graph + TF-IDF + OPTICS Clustering | FT |
| | LogM [108] | 2021 | L | DL | Sup, Unsup | CNN + Attention-based Bi-LSTM + K-means Clustering + Bag-of-words Model + TF-IDF + Siamese LSTM Network + Word2vec | C |

**Both.** LogM [108] offers unsupervised and supervised failure diagnosis techniques. The unsupervised approach is efficient but may be less accurate, while the supervised method performs better but requires annotated data. LogM first builds a knowledge base of root causes and abnormal events for the Hadoop platform. The unsupervised technique calculates cosine similarity between current and historical logs to identify probable root causes, while the supervised version uses a siamese LSTM network for similarity computation. Both methods aim to establish log-to-root cause mappings through pairwise similarity measurements. Additionally, LogM employs a CNN with attention-based Bi-LSTM to capture temporal dynamics in log sequences for failure prediction tasks.

*4.1.4 Summary.* Analyzing the above, most diagnostic techniques fall into three main approaches. First, category-based methods [59, 78, 79, 81, 83–85, 104, 107, 108] classify logs into different groups. Some of these techniques further diagnose categories, adding labels for common root causes and mitigation steps for online retrieval [59, 78, 79, 81, 104, 107, 108]. Others focus on distinguishing normal from abnormal log categories [84, 85] or use user-perceived metrics to assess category impact on performance degradation [83]. Second, correlation analysis techniques [80, 86, 87] employ weighted combinations or machine learning to evaluate feature correlations with system failures. Lastly, graph analysis methods [60, 88, 89] extract graph information from logs. ICWS'17 [88] builds a TCFG to mine execution flows and include interval information. HALO [89] creates an attribute hierarchy graph for root cause searches, while Swisslog [60] constructs an ID relation graph across distributed components.

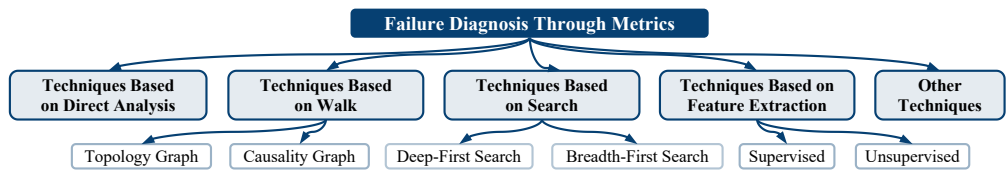

Fig. 5. Categories of failure diagnosis techniques through metrics.

Table 3 summarizes the surveyed failure diagnosis techniques through logs, providing information on the data required, the principles, the supervision techniques used, the core methods, and the diagnostic targets. It can be observed that the current root cause localization mainly focuses on the component-level, including log lines [78, 79, 84, 85, 88], events [82, 86, 87], impactful service system problems [83], with only a few techniques targeting the instance-level [59]. Regarding failure classification techniques [59, 104, 107], they cover the failure types compiled in Section 3.3.

In summary, logs are crucial for understanding system behavior at runtime, offering insights through natural language text, sequence information, and numerical values. They provide rich semantic information for analyzing performance issues and can be transformed into templates or events representing specific system behaviors through parsing [59, 60, 78, 79, 81–83, 85, 86, 88, 104, 107, 108]. Log sequences reflect system workflows, with changes often indicating potential failures [78, 79, 85, 88, 104]. Key variables in logs can also point to specific failure types. For instance, ICWS'17 [88] uses timestamps to extract time intervals between log prints for diagnosing latency anomalies. Notably, logs can be combined with user-perceived metrics for enhanced failure diagnosis [83, 86]. Log3C [83] clusters log sequences and analyzes their correlation with failure rates to identify problematic system behaviors. This approach replaces manual annotation with unsupervised correlation analysis, suggesting ways to reduce human involvement in failure diagnosis. Similarly, DISTALYZER [86] constructs graphs connecting log-extracted features with user-perceived metrics to identify strong dependencies, further advancing automated diagnostic techniques.

## 4.2 Failure Diagnosis Through Metrics

Metrics offer valuable insights into system resource usage and performance, providing fine-grained information to effectively characterize the operational state, particularly related to system resources. Failure diagnosis through metrics contributes to the observability and reliability of microservice systems, offering multi-dimensional, multi-granular, and multi-perspective techniques. Specifically, failure diagnosis utilizes collected metrics during operation, processing and analyzing them to establish diagnostic evidence by modeling correlations. As shown in 5, the techniques for failure diagnosis through metrics can be categorized into direct analysis, graph-based, search-based, feature-based, and other techniques.

*4.2.1 Techniques Based on Direct Analysis.* When a failure is detected by the front-end application of the system, these techniques primarily utilize traditional statistical methods to analyze the abnormal states of application monitoring metrics running in the back end to diagnose the root cause instances [61–66, 112], components[23, 65], or failure types [97, 106].

**Instance-level.** PAL [61] and FChain [62] propose two similar techniques to determine the root cause instances causing application-level performance anomalies. The core idea behind both approaches is to identify the root causes in distributed applications by extracting abnormal propagation patterns. They posit that performance anomalies manifest as significant changes in one or more system-level metrics, and these changes propagate from one instance to others, ultimately affecting the front-end application's service level objective (SLO). To find the root causes, PAL [61] initially applies a change-point detection algorithm [113] that combines cumulative sum (CUSUM) and Bootstrap to each collected system-level metrics after an exception occurs in the front-end application. This algorithm determines the start time of abnormal behavior. Then, PAL [61] orders the change points of different instances in chronological order to infer the abnormal propagation pattern. Typically, the first group of instances in the propagation pattern represents the root causes as they exhibit abnormal behavior earliest. In contrast to PAL [61], after obtaining the abnormal propagation pattern, FChain [62] filters out false positives and updates the pattern based on known instance dependencies, resulting in a more accurate diagnosis.

DBR [66] deploys an anomaly detector for each system instance and constructs a configuration profile for it based on historical data. During runtime, the instance anomaly detector checks if the current measurement deviates from the predicted value to identify and raise alerts for anomalies. When an anomaly is detected, information about the anomaly is sent to the anomaly correlation engine, which relies on domain knowledge captured from the current network and system configuration to analyze and identify the root causes. FluxRank [63] employs a root cause localization process comprising three steps, including change quantification, digest distillation, and digest ranking. The inputs for FluxRank [63] include the time of instance failure and all metrics. FluxRank [63] quantifies the degree of change for a large number of metrics using lightweight kernel density estimation (KDE). It then organizes all metrics into digests using DBSCAN to represent abnormal patterns in different modules. Subsequently, FluxRank [63] employs a learning-to-rank algorithm [112] based on feature extraction and logistic regression to automatically rank all digests based on their potential as root causes. Finally, operators mitigate the loss by triggering actions based on the digest ranking results. CloudPin [64] adopts a multidimensional algorithm with three sub-models to comprehensively analyze the diversity of anomalies in public cloud networks. In the prediction deviation dimension, CloudPin [64] designs a model based on the moving average algorithm. In the anomaly amplitude dimension, CloudPin [64] utilizes an improved model based on extreme value theory (EVT). In the shape similarity dimension, CloudPin [64] employs a set-based similarity model. It then designs a comprehensive sorting algorithm to generate the final ranking list of root cause instances, ensuring effective characterization of relative deviation based on the absolute deviation.

**Component-level.** CETS [23] introduces an event-time model that considers the duration of events, the relationships between events, and the correlations between events and various metrics. It transforms the problem into a dual-sample problem and uses the nearest neighbors method to explore the association between abnormal events and metrics, aiding in failure diagnosis.

**Both.** $\epsilon$-diagnosis [65] is a method that applies the two-sample hypothesis test to localize the root causes at the instance-level and component-level. It assumes that metrics with significant changes before and after a failure are more likely to be the root causes. $\epsilon$-diagnosis [65] first collects metrics on each container running the application and then extracts two equally sized time windows from it. One time window corresponds to the period when the front-end experiences anomalies, while the other corresponds to the period when the front-end operates normally. $\epsilon$-diagnosis [65] then calculates the similarity between the metrics within the two-time windows. If the similarity falls below a threshold, it indicates that the metric undergoes significant changes during the occurrence of anomalies, making it a candidate for the root cause. Finally, $\epsilon$-diagnosis [65] returns the abnormal components and their associated system instances as the set of potential root causes.

**Failure Type.** CloudPD [106] first collects various system-level metrics to generate data points, which are defined as sequences of moving average values over a fixed time interval. These data points serve as the basic input unit for various anomaly detection algorithms. Then, CloudPD [106] uses the KNN algorithm to learn the normal behavior in an online manner and compares the data points with the model to measure the deviation. During the failure classification stage, CloudPD [106] categorizes different failure types based on expert knowledge and generates failure signatures. It then compares the failure signatures with the existing ones in the database to determine the failure or generate new types. DBSherlock [97] is a heuristic algorithm that relies on scatter plots of visualized performance statistics. Users specify instances they consider abnormal, and the model automatically analyzes a large amount of historical statistical data to provide the most likely failure types and their corresponding confidence. DBSherlock [97] improves the accuracy of failure diagnosis by incorporating domain knowledge and employing optimization methods based on user feedback.

*4.2.2 Techniques Based on Walk.* Random Walk [114] is a mathematical statistical model where the process involves a sequence of trajectories, and each step of the walk is random. These techniques simulate the failure propagation among services using random walks and related variations. Each walk has the choice to stay at the current node or transition to a more abnormal node. After a certain number of iterations, a potential list of root causes is obtained based on the ranking of the walks. Nodes that are visited more frequently during the walks are more likely to be the root causes. Depending on the construction of the relationship graph that random walks rely on, there are two types based on topology graph [29–31, 34] and causality graph [32, 33, 35, 36, 45–48, 67, 68, 90, 91].

**Topology Graph.** MonitorRank [30] uses Hadoop to generate a call graph between system instances, and it utilizes random walks for root cause localization based on the call graph when instances experience anomalies. The core idea of this method is that metrics more correlated with the abnormal frontend nodes are more likely to be the root causes. To consider the interdependence between instances, three walk strategies are designed for simulating failure propagation. The transition probabilities between nodes are determined based on the metrics of the target nodes, the correlation with the frontend node, and the direction of the walk. Once an anomaly is detected in the frontend instance, MonitorRank [30] performs random walks based on the transition probabilities, and instances that are visited more frequently have a higher probability of being the root causes. ToN'18 [31] follows a similar approach to MonitorRank [30], suggesting that the correlation between virtual machine metrics and frontend instance response time can be used to measure the probability of virtual machines being the root causes. Specifically, ToN'18 [31] uses the tracing tool PreciseTracer [115] and the nova interface [116] to obtain the topology between virtual machines and constructs an anomaly propagation graph (APG) for each anomaly. Subsequently, random walk based on the APG is used to localize the root cause within the virtual machines.

The PageRank algorithm [117] initially developed as a method to calculate the importance of web pages on the internet, has also been widely applied in the domain of failure diagnosis in microservice systems. The basic idea is to define a random walk model, specifically a first-order Markov chain, on a directed graph that describes the behavior of visited nodes randomly. Under certain conditions, the probability of visiting each node converges to a steady distribution, and the steady probability values of each node represent their values, indicating their importance. MicroRCA [29] presents a failure diagnosis technique based on fine-grained feature engineering and personalized PageRank algorithm. Firstly, MicroRCA [29] detects anomalies in the system through an anomaly detection module. Once a performance anomaly is detected, the root cause analysis engine constructs an attribute graph with instances and nodes to represent the propagation path of the anomaly. Then, the root cause analysis engine extracts an anomaly subgraph from the attribute graph based on the detected anomaly. Finally, by calculating the anomaly scores of nodes and the correlations between nodes, MicroRCA assigns node weights and edge weights on the anomaly subgraph and uses the personalized PageRank algorithm to infer the instances that are most likely to cause the anomaly. Building upon the work of MicroRCA [29], ICSOC'20 [34] further considers the root cause components that lead to frontend instance failures. Firstly, it constructs an instance dependency graph to discover candidate root cause instances. Then, it takes into account the abnormality levels of the instance metrics before and after the failure. To achieve this, ICSOC'20 [34] trains an autoencoder to learn the patterns of these instance metrics under normal circumstances. After obtaining the candidate root cause instances, the corresponding autoencoder is used to reconstruct the metrics after the occurrence of anomalies. If the reconstruction results are poor, it indicates that these components and their corresponding instances are more likely to be the underlying causes.

**Causality Graph.** It is a directed acyclic graph that captures the dependency relationships between instances in the failure propagation. CloudRanger [45], MS-Rank [46], and AutoMAP [47] utilize the PC algorithm [110] to analyze performance metrics and build dynamic causality graphs between instances. They then employ the random walk algorithm similar to MonitorRank [30] to determine the root cause instances leading to failures. MS-Rank [46] improves upon CloudRanger [45] by considering multiple metrics from different instances when constructing the causality graph. It also introduces a feedback mechanism that allows the model to update the importance levels of different instance metrics based on feedback from operators, improving the accuracy of future diagnostic results. AutoMAP [47] introduces the concept of addition and subtraction operations on graphs, where the relationship graphs constructed under normal conditions are summed to display the instance linkage. This helps reduce noise interference. Additionally, AutoMAP [47] subtracts the abnormal behavior graph obtained through causal analysis from the normal graph to retain a few instances that are highly relevant to the anomaly, making the abnormal information in the graph more prominent and improving the accuracy of subsequent failure diagnosis.

Other techniques [32, 33, 48, 67, 90] utilize different methods to construct causality graphs. MicroCause [90] designs a simple yet effective path condition time series (PCTS) algorithm to capture the time lag characteristics in the failure propagation between metrics. It then uses a novel temporal cause-oriented random walk (TCORW) algorithm that considers causal relationships, temporal order, and priority information of the metrics to rank the root cause components. ServiceRank [48] treats the cloud-native system as a "black box" and constructs an influence graph by extracting causal relationships between instances using the PC algorithm [110] without any predefined topology. To enhance the reliability and availability of services, operators have developed various design patterns to provide failure tolerance capabilities. However, these patterns change the way failures propagate, rendering traditional diagnostic methods ineffective. To address this issue, ServiceRank [48] proposes a correlation calibration mechanism to eliminate the impact of design patterns on failure diagnosis. Finally, ServiceRank [48] designs a heuristic investigation algorithm based on the second-order random walk to localize the root causes. Similar techniques include FRL-MFPG [67], which proposes a microservice failure propagation graph construction method based on failure correlation (MFPG-FC) to study the failure propagation patterns of instances in a system. In response to the limitations of inferring root causes based on correlation calculation and to avoid being trapped in a low abnormality area, FRL-MFPG [67] designs a random walk algorithm with forward, backward, and stay access, which can accurately localize. REASON [32] addresses the problem of localizing root cause instances in complex systems with interdependent network structures. It proposes a method based on hierarchical graph neural networks (GNN) to construct causality graphs that consider both intra-level and inter-level non-linear causal relations. Then, REASON [32] uses random walk with restarts (RWR) to model the network propagation of system failures to identify potential instances. CORAL [33] designs an online technique that can automatically trigger the failure diagnosis process and incrementally update the model, addressing the inefficiency of offline techniques. CORAL [33] first detects the trigger point for failure diagnosis using metrics. If triggered, it initiates the incremental causality graph learning process using the variational graph autoencoder (VGAE). In the incremental causality graph learning phase, each data batch is used to decouple the state-invariant and state-dependent information to incrementally update the previous causality graph. Finally, similar to the other techniques, CORAL [33] uses RWR based on the learned causality graph to simulate failure propagation. When the learned causality graph and the list of root cause instances converge, the final diagnostic result is obtained.

The core idea of LOUD [68], MicroDiag [35], and CauseRank [91] is based on constructing a causality graph and using standard or personalized PageRank algorithms to localize the root causes. LOUD [68] is based on the assumption that failure resources will generate increasingly strong correlations with time during failure execution. It first trains a model that captures the normal behavior of the system and only retains the anomalous indicators during the online phase. It then deduces the propagation graph based on the causality graph generated during training. Finally, graph centrality algorithms such as PageRank are used. MicroDiag [35] collects the call relationships and metrics of different components and uses an unsupervised algorithm named distance-based clustering BIRCH for anomaly detection. It analyzes the collected user-perceived and system-level metrics, obtains service dependency relationships and deployment information, and constructs a component dependency graph. Further, it uses a structural causal model (SCM) [118] to infer causal relationships and construct a metrics causality graph. Finally, the PageRank algorithm is used to localize the root cause components. CauseRank [91] is a failure diagnosis technique at the metric group level. Firstly, CauseRank [91] groups the metrics in the system based on their respective modules to reduce the complexity caused by high-dimensional metrics. Secondly, CauseRank [91] calculates the volatility of each metric to filter out candidate metric groups that are related to failures. Then, a group-based greedy equivalent search (G-GES) algorithm is used to construct a temporary causality graph between the candidate metric groups. Finally, CauseRank [91] designs a causal-oriented personalized PageRank (COPP) algorithm to score and rank the candidate component groups, thereby determining the most likely root causes.

In large language models like ChatGPT, human-in-the-loop training has been proven to be effective in improving model performance. Inspired by this, HRLHF [36] combines human feedback from experts familiar with system architecture and diagnostic experience to improve the accuracy of constructing the service dependency graph during the process of identifying system causal relationships. Additionally, HRLHF [36] enhances CausalRCA [119] by transforming static causality graphs into window causality graphs, which incorporate the characteristics of Markov and Granger causality. This ultimately leads to more accurate and robust results.

### 4.2.3 *Techniques Based on Search.*
Techniques based on search typically rely on information such as metrics, network traffic, and invocation relationships between system services or instances to construct a DAG with metrics or system instances as nodes and causal relationships as edges. The entire causality graph is then traversed using strategies like deep-first search (DFS) [25, 37, 69] or breadth-first search (BFS) [38, 70] to find potential root causes.

**Deep-First Search.** CauseInfer [37], IPCCC'16 [25], and Microscope [69], all based on constructed causality graphs, use the DFS strategy to identify the root causes of performance anomalies in application front-ends. CauseInfer [37] proposes a two-layered hierarchical causality graph construction method. Coarse-grained graphs are constructed based on the lag correlation of the sending traffic between two instances, while fine-grained graphs between internal components are built using the PC algorithm [110]. Then, CauseInfer [37] employs the DFS strategy to traverse the entire causality graph and prioritize and infer potential components according to anomaly scores. Specifically, CauseInfer [37] starts with the SLO metrics of the application front-end and recursively visits nodes in the causality graph. For each visited node, it checks if the dependent nodes exhibit abnormal behavior. If not, the currently visited node is considered the root cause.

IPCCC'16 [25] improves the accuracy by incorporating operational personnel's feedback knowledge into the process of using DFS to search on the causality graph. IPCCC'16 [25] first extracts the causality graph using data mining techniques. Then, when a failure occurs, a greedy DFS algorithm is applied to generate candidate root causes. Operational personnel validate the candidate's root causes and provide feedback indicating correctness or incorrectness. Finally, a supervised learning algorithm is used to update the causality graph based on the feedback, enhancing the accuracy of DFS in outputting candidate root causes.

Microscope [69] takes into account both communicating and non-communicating dependency relations between two service instances to construct the causality graph. In some cases, instances may not have communication dependencies but may share computational resources on the same physical machine, leading to potential failure propagation. Microscope [69] uses a parallelized PC algorithm [110] to learn causal relationships arising from such shared resources. After constructing the causality graph, Microscope [69] starts from anomalous front-end instance nodes and uses the DFS strategy to traverse the graph, adding anomalous instances with all neighboring nodes in a normal state to the candidate list. Finally, Microscope [69] calculates the Pearson correlation coefficient between the front-end instance and each candidate instance's SLO metric as the basis for ranking the root causes.

**Breadth-First Search.** AS'20 [38] and DyCause [70] both based on constructed causality graphs, use the BFS strategy to identify the root causes of performance anomalies in application front-ends. AS'20 [38] efficiently constructs causality graphs between metrics by utilizing an operation and maintenance knowledge graph. It assigns weights to each edge based on the Pearson correlation coefficient between two sequences, where the sequence represents the value changes for the node. When an abnormal metric is detected, AS'20 [38] applies the BFS strategy starting from the abnormal metric to find all possible paths. Finally, AS'20 [38] ranks the paths based on the sum of edge weights. If the same, it prioritizes the shorter path as the root cause.

DyCause [70] introduces a technique using sliding windows and crowdsourcing graph fusing. Firstly, DyCause [70] examines Granger causal intervals with sliding windows to construct local dependency graphs. These graphs represent the extent of influence an instance has on the front-end application and other instances. To leverage collective wisdom, DyCause [70] designs a crowdsourcing graph fusing scheme that merges the local dependency graphs from different instances into an optimized dependency graph. Finally, based on the optimized dependency graph, DyCause [70] performs BFS in reverse, constructing abnormal propagation paths and generating a ranked list of root cause instances.

*4.2.4 Techniques Based on Feature Extraction.* Techniques based on feature extraction refer to the use of machine learning techniques to automatically learn performance feature models from metrics. These models can describe the performance behavior of the instances or components under normal operating conditions and different failure states. When new metrics are collected, the trained feature models are used to analyze the data and complete the failure diagnosis task. The feature model maps the original data states to a new feature space, which may better meet the requirements of specified tasks compared to the original space. Typically, the mapped data is fed into commonly used machine learning algorithms such as clustering or classification algorithms to obtain corresponding models. Based on the different supervision methods, feature-based techniques can be divided into supervised [43, 92, 101, 120] and unsupervised techniques [44, 93, 98, 103].

**Supervised.** ISSRE'16 [101] utilizes monitoring metrics from runtime system instances that reflect the system-level resource status to detect abnormal system behavior. It then trains a random forest model for anomaly behavior classification. PatternMatcher [43] takes into account the varying levels of attention that operators have for different abnormal patterns of metrics. PatternMatcher [43] first leverages two-sample hypothesis tests for coarse-grained anomaly detection on all metrics to quickly and accurately filter out metrics that do not show any abnormal changes during the occurrence of failure, narrowing down the search space. Then, PatternMatcher [43] uses a one-dimensional CNN (1-D CNN) to extract features from labeled time series data and trains an anomaly pattern classification model [121] to further analyze the abnormal patterns of the metrics. The aim is to filter out the abnormal patterns that engineers are not concerned about, thereby improving localization accuracy and providing engineers with more comprehensive information. Finally, PatternMatcher [43] employs a ranking strategy that considers both the degree of metric anomaly and the importance of abnormal patterns to automatically rank the abnormal metrics, allowing operators to inspect suspicious components based on the ranking list.

Due to the natural graph structure that facilitates modeling the topology of the system, Arvalus [120] and DéjàVu[92] use GNN to learn features and patterns from graph structure data, showing good performance in failure diagnosis. Arvalus [120] considers the dependencies and failure propagation relationships among cloud-native system instances. It first transforms the metric subseries of system instances into feature vectors of graph nodes and then uses graph convolution operations to learn the features and weights of system instance nodes and edges. Finally, it combines the node features of system instances with dependency relationship features and performs failure classification through feature transformations using linear and softmax layers. DéjàVu[92] first captures the temporal information of metrics and the correlations between metrics. DéjàVu[92] trains a feature extractor for each failure type to map the instances of the same type to vectors of the same dimension. To model failure propagation, the feature aggregator utilizes the attention mechanism to aggregate the structural information of the failure dependency graph (FDG) into one aggregated feature. The faulty instances of the same type share a feature extractor, and all faulty instances share a feature aggregator. Finally, in conjunction with FDG, GNN is used to diagnose the root cause instances and specific failure types.

**Unsupervised.** Fingerprint [103] aims to automatically diagnose failure types by identifying recurring behaviors that may occur due to misconceptions of the root cause, delayed deployment fixes, or sudden actions caused by high utilization. Fingerprint [103] first captures the state of each metric and calculates percentiles to identify abnormal behavior. It then succinctly summarizes the collected subset of metrics that can best differentiate between different failures. Based on the historical percentiles of each metric, the current value is described as hot, cold, or normal, indicating an up, down, or normal value, respectively. Finally, feature encoding is obtained, and the failure type is matched using Euclidean distance. iSQUAD [98] consists of offline analysis and training phase, and online diagnosis and update phase. In the offline phase, the collected metrics are first extracted for abnormal features, including peak rising or falling and level shifting. Then, iSQUAD [98] applies dependency cleansing based on association rule learning between pairs of metrics. Afterward, iSQUAD [98] utilizes type-oriented pattern integration clustering (TOPIC) to obtain clustering clusters. To quickly find the nearest neighbors, TOPIC uses the KD-tree method. Finally, iSQUAD [98] selects important metrics and representative failure features through the Bayesian case model (BCM) to represent the entire cluster, and this extracted key information is handed over to operations personnel to label the failure types. BCM explains clustering or classification results using representative samples and provides model interpretability. In the online diagnosis phase, the same steps are followed to extract features, match them with pre-trained typical features, and diagnose the failure type based on the feature with the highest similarity. TS-InvarNet [93] is based on the assumption of stable relationships between metrics and aims to mine and interpret state changes of invariants for root cause localization. TS-InvarNet [93] first uses the hierarchical DBSCAN (HDBSCAN) algorithm to eliminate duplicate and redundant invariants to accelerate the construction of the invariant network. TS-InvarNet [93] learns global and local dependency relationships through the tempo-spatial model. Then, TS-InvarNet [93] detects anomalies based on the evolution of the global invariant network and performs root cause localization based on the interpretation of changes in local invariants.

RootCLAM [44] not only achieves root cause localization from a causal perspective but also focuses on anomaly mitigation. RootCLAM [44] manages data generation using SCM [118] and considers root causes as external interventions on specific features. Specifically, RootCLAM [44] first uses a deep support vector data description-based (Deep SVDD) anomaly detection algorithm to obtain abnormal features and trains a variational causal graph autoencoder (VACA) on normal data to capture the normal data distribution. When obtaining abnormal features, RootCLAM [44] deduces the hidden variables for each abnormal feature and calculates the cumulative probability of each exogenous variable based on its fit to the normal data distribution. Finally, a comparison between each cumulative probability and a predefined threshold determines the variable sets that serve as the root causes for that abnormal feature.

*4.2.5 Other Techniques.* The remaining techniques can be categorized into service-level [49], instance-level [24, 71, 72, 77], and component-level [40–42, 94] based on their finest granularity of failure diagnosis.

**Service-level.** FacGraph [49] proposes a frequent pattern mining algorithm on an anomaly correlation graph to discover root cause instances. FacGraph [49] consists of three steps. First, it constructs a causal relationship graph based on delay and throughput metrics using the PC algorithm [110] and d-separation [65]. Then, FacGraph [49] applies the breadth-first ordered string (BFOS) for frequent graph mining (FSM) on the causality graph and scores the subgraphs. Finally, FacGraph [49] filters out high-scoring subgraphs and returns the corresponding instance sets of leaf nodes in the subgraphs as the set of root cause instances. FacGraph [49] also develops a distributed version that utilizes parallel computing to accelerate the FSM process.

**Instance-level.** NetMedic [71] periodically captures the state of each system instance as multi-variable vectors and stores the states. However, the number and meaning of state variables for each instance may vary. NetMedic [71] captures instance status through dependency templates to generate dependency graphs. Then, NetMedic [71] calculates the anomaly scores for each instance based on historical data and the weights of edges in the dependency graph to rank the root cause instances. It is worth mentioning that NetMedic [71] exhibits good scalability and can be improved to achieve finer-grained root cause localization.

CRD [72] discovers multiple failure propagations occurring simultaneously in different node clusters, collectively defining the system's state. Conventional root cause localization methods typically assume a single failure propagating in the network, and to address this, a multi-root cause technique is proposed. This method consists of two stages. In the first stage, CRD [72] proposes a joint clustering model that utilizes complementary information from the invariant network and the broken network to identify and rank clusters in the invariant network. In the second stage, CRD [72] designs a low-rank network diffusion model to backtrack causal anomalies in the impaired node clusters identified in the first stage. CRD [72] can handle parallel and localized failure propagations in different clusters, making it suitable for scenarios with multiple causal anomalies.

Grano [24] presents a graph-based interactive root cause analysis method composed of the anomaly detection layer, the anomaly graph layer, and the application layer. Initially, users extract metric time series data from the indicator database to obtain detection events in the anomaly detection layer using the corresponding detection models. The anomaly graph layer serves as the fusion point, where the detection events, application events, and topology structure are fused to construct the anomaly graph, which is then stored in the database. Finally, in the application layer, the anomaly graph algorithm is used to obtain root cause relevance scores for system components.

JSS'20 [77] proposes a graph comparison-based failure diagnosis framework. JSS'20 [77] first uses existing anomaly detection techniques to detect anomalous instances and constructs anomalous subgraphs, which are graph structures containing nodes adjacent to the anomalous instances. The anomalous subgraphs are then compared with previously excluded anomalous graph patterns by the operations personnel. If the similarity exceeds a threshold, the anomalous subgraphs are considered to contain potential root cause instances. If multiple root cause instances are detected, they are ranked based on the similarity between the corresponding anomalous graph patterns and anomalous subgraphs.

**Component-level.** Sieve [40] addresses challenges faced by large-scale distributed systems, including a high volume of monitoring metrics and the complexity of combining system instance dependencies and metrics. Firstly, Sieve [40] reduces the dimensionality of metrics by filtering out unimportant indicators using a clustering-based centroid preservation method. Then, Sieve [40] performs Granger causality tests on system instances with existing call relations, employing a predictive-causality model to infer the dependency relationships between system instances. Finally, Sieve [40] compares the differences between the dependency graphs during normal and abnormal states.

DLA [41] models the topological structure of system instances using a hierarchical hidden Markov model (HHMM) and calculates the most likely paths in the HHMM that lead to observed anomalies on instances. These paths are then used to deduce the most probable root cause components that could cause the observed anomalies.

ExplainIt [94] presents a declarative, unsupervised root cause analysis engine. ExplainIt [94] utilizes probabilistic graphical models (PGMs) for causal inference, enabling precise localization in large-scale metric databases.

CIRCA [42] formulates the task of online instance-level root cause localization as a new causal inference task called intervention recognition. CIRCA [42] first proposes a method to construct a causal Bayesian network (CBN) based on the system architecture. Then, CIRCA [42] employs regression-based hypothesis testing and descendant adjustment methods to infer the root cause components in the network.

*4.2.6 Summary.* Researchers have combined metrics with advanced algorithms for failure diagnosis, yielding promising results due to metrics' ability to directly reflect system resource issues and represent system characteristics. Some techniques enhance diagnostic performance by leveraging system topology [29–38, 40–44, 77] or event information [23, 25]. Table 4 categorizes the surveyed failure diagnosis techniques using metrics, indicating the required data, underlying principles, supervision techniques, and core algorithms employed.

Regarding root cause localization techniques [23–25, 29–38, 40–49, 61–72, 77, 90, 91, 93, 94], various approaches target different systems and localize root causes at varying granularities. These include service-level (front-end, back-end, and APIs), instance-level (pods, containers, hosts, processes, VMs, and servers), and component-level (metrics[35, 42–44, 65, 68, 90–92, 94], code[94], and events [23, 25, 94]). Except for Arvalus[120], failure classification techniques [92, 97, 98, 101, 103, 106] address hardware, software, and network failures as outlined in Section 3.3.

Direct analysis techniques fall into two main categories: pattern mining and correlation analysis. Early research focused on mining anomaly propagation patterns [61, 62, 66], identifying the initial instances in these patterns as root causes. Later work explored more comprehensive anomalous state information, such as metric change degrees [63]. Correlation analysis identifies components most closely linked to anomalous events or metrics, examining correlations between events and metrics [23] or among anomalous metrics [65]. Recent methods like CloudPin[64] combine both approaches in their design through three sub-models.

Graph theory algorithms have proven highly effective in failure diagnosis. Researchers have constructed system topology [25, 29–31, 34] or causality graphs [32, 33, 35–38, 40–42, 44–49, 67–72, 77, 90–94, 101, 120] by exploring invocation relationships between system services, instances, or components. From CloudRanger [45] to HRHLF

Table 4. Summary of surveyed failure diagnosis techniques through metrics, based on their *category*, the publication *year*, the *data* (*e.g.*, M for metrics, E for events, and TP for topology) they use, the *principle* (*e.g.*, R for techniques based on rule, ML for techniques based on machine learning, and DL for techniques based on deep learning) and *supervision* (*e.g.*, Sup for supervised techniques, and Unsup for unsupervised techniques) type, the core *method*, and the diagnostic *target* (*e.g.*, S for service-level localization, I for instance-level localization, C for component-level localization, and FT for failure types).

| Category | Technique | Year | Data | Principle | Supervision | Core Method | Target |
|---|---|---|---|---|---|---|---|
| Direct Analysis | PAL [61] | 2011 | M | R | Unsup | CUSUM + Bootstrap [113] +Propagation Pattern | I |
| | FChain [62] | 2013 | M | R, ML | Unsup | CUSUM + Bootstrap [113] +Propagation Pattern | I |
| | DBR [66] | 2016 | M | ML | Unsup | K-means Clustering + Anomaly Propagation Graphs Pattern | I |
| | FluxRank [63] | 2019 | M | ML | Unsup | DBSCAN + Learning-to-rank[112] | I |
| | CloudPin [64] | 2021 | M | ML | Unsup | Moving Average + EVT + Set-based Similarity | I |
| | CETS [23] | 2014 | M, E | ML | Sup | Nearest Neighbors Method | I |
| | e-diagnosis [65] | 2019 | M | R | Unsup | Two-sample Test Algorithm + e-statistics Test | I, C |
| | CloudPD [106] | 2013 | M | R, ML | Unsup | KNN | FT |
| | DBSherlock [97] | 2016 | M | ML | Unsup | Optimization Methods based on User Feedback | FT |
| Walk-based | MonitorRank [30] | 2013 | M, TP | ML | Unsup | Personalized PageRank + Pseudo-anomaly Clustering Algorithm | S |
| | ToN'18 [31] | 2018 | M, TP | R, ML | Unsup | APG + Random Walk | I |
| | MicroRCA [29] | 2020 | M, TP | R, ML | Unsup | Anomalous Subgraph + Personalized PageRank | I |
| | ICSOC'20 [34] | 2020 | M, TP | ML | Unsup | Personalized PageRank + Auto-encoder | S, C |
| | CloudRanger [45] | 2018 | M | R, ML | Unsup | PC [110] + Heuristic Investigation Algorithm based on Second-order Random Walk | S |
| | MS-Rank [46] | 2019 | M | R, ML | Unsup | PC [110] + Random Walk | S |
| | AutoMAP [47] | 2020 | M | R, ML | Unsup | PC [110] + Heuristic Random Walk | S |
| | MicroCause [90] | 2020 | M | ML | Unsup | PCTS + TCORW | C |
| | ServiceRank [48] | 2021 | M | R, ML | Unsup | PC [110] + Heuristic Investigation Algorithm based on Second-order Random Walk | S |
| | FRL-MFFG [67] | 2023 | M | R, ML | Unsup | MFFG-FC + Random Walk | I |
| | REASON [32] | 2023 | M, TP | DL | Unsup | Hierarchical GNN + RWR | I |
| | CORAL [33] | 2023 | M, TP | DL | Unsup | VGAE + RWR | I |
| | LOUD [68] | 2018 | M | ML | Unsup | Propagation Graph + PageRank | C |
| | MicroDiag [35] | 2021 | M | ML | Unsup | SCM [118] + PageRank | C |
| | CauseRank [91] | 2022 | M | ML | Unsup | G-GES + COPP | C |
| | HRLHF [36] | 2023 | M, TP | ML, DL | Sup | PC [110] + Human Feedback + Learned Reward Function | S |
| Search-based | CauseInfer [37] | 2014 | M, TP | ML | Unsup | PC [110] + DFS | C |
| | IPCCC'16 [28] | 2016 | M, E | ML | Sup | Greedy DFS Algorithm + Random Forest | I, C |
| | Microscope [69] | 2018 | M | ML | Unsup | Parallelized PC + DFS | I, C |
| | AS'20 [38] | 2020 | M, TP | R | Unsup | Optimized PC based on Knowledge Graph + BFS | I, C |
| | DyCause [70] | 2021 | M | R | Unsup | Crowdsourcing Graph Fusing + BFS | C |
| Feature-based | ISSRE'16 [101] | 2016 | M | ML | Sup | Random Forest | FT |
| | PatternMatcher [43] | 2021 | M, TP | DL | Sup | Two-sample Hypothesis Test + 1-D CNN + Multi-layer Perception | C |
| | Arvalus [120] | 2021 | M | DL | Sup | BIRCH [122] + Graph Convolutional Neural Networks | FT |
| | DéjàVu[92] | 2022 | M, TP | DL | Sup | FDG + GNN | C, FT |
| | Fingerprint [103] | 2014 | M | ML | Unsup | Recognition + Hot and Cold Metric Quantiles + Crisis Fingerprint Comparsion | FT |
| | iSQUAD [98] | 2020 | M | ML, DL | Unsup | TOPIC + BCM | FT |
| | TS-InvarNet [93] | 2022 | M | ML | Unsup | HDBSCAN + Granger Causality Test | C |
| | RootCLAM [44] | 2023 | M, TP | DL | Unsup | SCM [118] + Deep SVDD Anomaly Detection Algorithm + VACA | C |
| Other | FacGraph [49] | 2018 | M | R | Sup | PC [110] + D-separation + BFOS + FSM | S |
| | NetMedic [71] | 2009 | M | ML | Unsup | Dependency Graph | I |
| | CRD [72] | 2017 | M | ML | Unsup | Doubly Stochastic Matrix Decomposition | I |
| | Grano [24] | 2019 | M, E, TP | R, ML | Unsup | Unified Anomaly Graph + Propagation Algorithm | I |
| | JSS'20 [77] | 2020 | M, TP | ML | Unsup | Anomalous Subgraph + Graph Comparison | I |
| | Sieve [40] | 2017 | M, TP | R | Unsup | K-shape Clustering + Predictive-causality Model | I, C |
| | DLA [41] | 2019 | M | DL | Unsup | Baum-Welch Algorithm + HHMM | I, C |
| | ExplainIt [94] | 2019 | M | ML | Unsup | CBN + Hypothesis Ranking | C |
| | CIRCA [42] | 2022 | M, TP | DL | Unsup | SCM [118] + CBN | C |

[36], most techniques use the PC algorithm [110] to construct causality graphs. Microscope [69] and AS'20 [38] have improved this algorithm for better system adaptation. In these graphs, nodes represent system elements, and edges depict invocation or causal relationships. Various graph theory algorithms are employed in different system environments. Random walk is the most common technique [29–36, 45–48, 67, 68, 90, 91], where each step involves staying at the current node or moving to a more anomalous one. MonitorRank [30] proposed three random walk strategies, influencing subsequent studies [31, 45–47]. Personalized PageRank algorithms [29, 34] are also widely used. Some techniques use DFS [25, 37, 69] or BFS [38, 70] to traverse the graph from an anomalous node.

With advancements in AI, feature-based techniques [43, 92, 93, 98, 101, 120] extract metrics and input them into various models to describe performance under different states. CNN [101] and GNN [44, 92, 120] are used for feature extraction and pattern mining, with GNN and relationship graphs yielding particularly good results. Other techniques [24, 40–42, 49, 71, 72, 77, 94] employ various inference analysis methods, such as calculating scores [49], comparing similarities [77], or building causal analysis models [40–42, 94] for failure diagnosis.

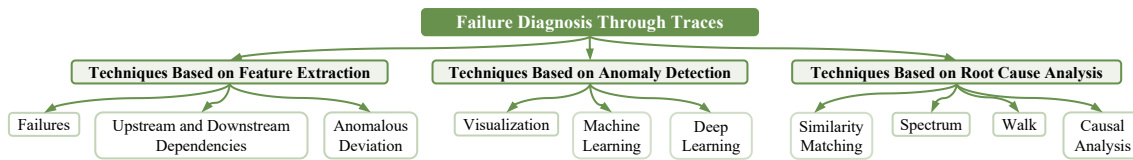

Fig. 6. Categories of failure diagnosis techniques through traces.

## 4.3 Failure Diagnosis Through Traces

Before the popularization of traces and the development of end-to-end trace generation and collection systems, software systems have used runtime path [123] and client request traces [99, 124] to record interaction information during the operation of distributed systems. Some failure diagnosis techniques through them has gradually validated their value. For example, Pinpoint [99] clusters instances based on the observation that instances involving root causes are likely to result in failed user requests. It forms vectors based on the occurrence of each instance in various user requests and vectors representing the failure situations of all user requests in the system, allowing the identification of sets of root cause instances. PBAA [123] designs a software system based on the runtime path to manage failure detection and diagnosis, failure impact analysis, and system evolution understanding. Chen et al. [124] train decision trees on the request traces. By examining the paths leading to failure-predicting leaf nodes, operators can gain insights into potential root causes.

With further study, an increasing number of end-to-end trace generation and collection systems have been designed. These systems typically include clients for collecting and sending spans, collectors for gathering spans, backend storage for persistent data storage, and APIs and UI dashboards for users to query traces. Google's Dapper [125] defines tracepoints as marked timestamps and four key activities on each span, including *server send*, *client receive*, *client send*, and *server receive*, representing server sending a request, client receiving a request, client sending a response, and server receiving a response. The implementation and adoption of these systems, such as Magpie [126], X-Trace [127], Dapper [125], and Startdust [128], have demonstrated their value in production systems.

The process of failure diagnosis through traces generally consists of three steps: (1) Feature Extraction [17, 52, 53, 57, 58, 74, 129]. Extracting features from the collected traces. (2) Anomaly Detection [1, 20, 22, 74, 76]. Some methods perform anomaly detection on traces before failure diagnosis, specifically distinguishing normal and abnormal traces. (3) Root Cause Analysis [21, 25, 27, 54–56, 74, 75, 105, 130]. Combining the extracted features, the normality of traces, and other information to complete failure diagnosis. As shown in Figure 6, different studies in this domain also focuses on improving and innovating in a specific step of the process.

*4.3.1 Techniques Based on Feature Extraction.* Feature extraction refers to obtaining information related to failure diagnosis from the traces. Based on them, statistical, machine learning, deep learning, and other methods can be used to complete failure diagnosis tasks. We summarize three common types: features related to failures [17], upstream and downstream dependencies [52, 53, 74], and anomalous deviation [57, 58, 129].

**Failures.** MEPEL [17] summarizes a set of features that reflect the dynamic environment and interactions of microservices, describing the state of the microservice system from the perspectives of configuration, resources, instances, and interactions. Configuration features reflect the environmental configuration of microservice instances. Resource features reflect the consumption of resources by microservice instances and their deployment nodes (*e.g.*, memory and CPU consumption). Instance features reflect the deployment status of microservice instances and their participation in the current trace instance. Interaction features reflect the interaction status with other microservices, especially asynchronous interactions. Based on the extracted features, MEPEL [17] trains a series of models, including random forest, KNN, multilayer perceptron (MLP), for latent error prediction, faulty microservice prediction, failure type prediction, and microservice status prediction.

**Upstream and Downstream Dependencies.** RanCorr [53] defines an aggregation method for anomaly scores. The core idea is that if the anomaly score of the caller is greater than the anomaly score of the operation under consideration in a calling dependency graph, then the operation is likely to be the root cause because the dependent operations exhibit significant anomalies. If the maximum anomaly score of the directly connected callee is greater than the anomaly score of the current operation, it means that the anomaly of this operation likely originates from the propagation of another operation it depends on. RanCorr [53] considers the correlation between root cause nodes and failure propagation by aggregating the anomaly scores of upstream and downstream nodes. ModelCoder [74] uses traces to construct the deployment graph and the service dependency graph and divides anomalous service nodes into explicit and implicit nodes. The explicit nodes are the initial nodes of anomalous calls, and the implicit nodes are the response nodes of anomalous calls. Based on the explicit and implicit nodes, including the target node itself, its child nodes, its parent nodes, and its bidirectional nodes, ModelCoder [74] proposes node feature encoding, which calculates the similarity between service nodes with unknown root cause and service nodes with known root cause in a standard code storage, matching the failure type and localizing the root cause. However, ModelCoder [74] requires a standard storage that covers as many failure types as possible, and for new failures that have never been encountered before, it may produce incorrect diagnostic results. MicroHECL [52] predefines three types of anomalies: performance, reliability, and traffic. Based on the quality metric and propagation direction of each type of anomaly, MicroHECL [52] extracts specific statistical features and applies specific machine learning methods for detection. If the upstream and downstream relationships of the initial anomalous service are consistent with the failure propagation direction detected, it will iteratively extend to establish a propagation chain by backtracking along the anomaly propagation direction from the starting point. When no new chains can be established, the candidate root cause is selected from the end of the failure propagation chain. In addition, MicroHECL [52] combines the upstream and downstream dependency relationships in the graph with the Pearson correlation coefficient. For example, during the construction of the failure propagation chain, pruning is performed based on the correlation coefficient of the quality metrics of two successive service calls. The correlation between the change trend of the initial anomalous service and the Pearson correlation coefficient of the candidate root cause node is measured to provide the ranking of the root cause.

**Anomalous Deviation.** It is a common failure feature that quantifies the difference between abnormal and normal states. CloudDiag [129] utilizes matrix decomposition to extract this feature. The method first calculates the coefficient of variation [131] for the same class of requests with the same call tree. If it exceeds a given threshold, it indicates that there is a significant deviation within this class of requests, suggesting the presence of anomalies. Then, each method's execution time in each abnormal request is combined to form a matrix. Each column of the matrix represents the time vector of the corresponding method, and each row represents a specific request. Intuitively, requests with similar call trees have similar execution times, meaning that the rows of the matrix are correlated. The robust PCA (RPCA) [132] algorithm can decompose this matrix into a low-rank matrix with non-corrupted columns and a sparse matrix with a few nonzero corrupted columns. The former represents the normal state of the system, while the latter represents the degree of anomalous deviation. If the angle between the original columns and the corrupted columns exceeds a given threshold, the column corresponding to the method is determined to be anomalous. By counting the number of anomalies for each method across different types of requests, the final ranking of root cause methods can be obtained. WinG [58] utilizes the distance obtained from the dynamic time warping (DTW) algorithm to characterize the anomalous deviation. First, WinG [58] collects and computes the average latency of each invocation pair over a one-minute period to generate a feature vector. Then, using the DTW algorithm, WinG [58] calculates the distance between the feature vectors of the current period and the normal period, serving as a quantified score for the degree of anomaly. Grouping by microservice type, the maximum degree of anomaly relative to other degrees of anomaly within each group is computed. This ratio is used as the basis for ranking the root causes of microservices. Finally, considering the frequency of microservice anomalies occurring over a period of time, WinG [58] filters out potential false positives and recommends microservices with longer durations of anomalies as the root causes. In contrast to WinG [58], TraceNet [57] uses an abnormal score based on the 3-sigma principle to characterize the anomalous deviation. TraceNet [57] aggregates the average latency and standard deviation at the operation level of the trace. The ratio of the latency of the target trace relative to the standard deviation away from the mean is considered an anomaly feature. Then, the operations involved in each microservice instance are divided into upstream and downstream impacts. The weighted sum of these impacts yields an anomaly score for the microservice instance. By combining the proportion of abnormal nodes in the upstream and downstream of the microservice instance, the final ranking of root causes is determined.

*4.3.2 Techniques Based on Anomaly Detection.* Accurately determining the abnormality of traces is fundamental to some failure diagnosis techniques. Abnormal and rare traces often provide strong indications of the root cause. In this section, we categorize common trace anomaly detection techniques based on visualization [1], machine learning [20], and deep learning [22, 76].

**Visualization.** These techniques determine possible root causes by comparing traces with trace visualization tools. GMTA [1] implements a graph-based trace analysis system that supports various functionalities, such as visualizing the dependency graph between microservices, analyzing changes in microservice behavior, detecting performance issues, and pinpointing root causes. Operators can use this system to compare business flows before and after failures occur, obtain EP chains, narrow down the scope of root causes to specific paths or operations, and further localize the root cause. GMTA [1] focuses on system design and requires more human effort compared to automated anomaly detection.

**Machine Learning.** MicroSketch [20] groups the collected trace spans based on the types of upstream and downstream instances. It then constructs an indefinite-length vector for each group by taking the handling time quantiles. During the anomaly detection phase, MicroSketch [20] maintains a robust random cut forest (RRCF). The specific technique is as follows: for each system state vector to be added, RRCF randomly selects a dimension and cuts the vector space at a randomly chosen value in that dimension, dividing the space into two parts. This process is repeated recursively, resulting in each node in the tree dividing the state vectors in its subtree based on a dimension's size. After the graph is constructed, the difference in sizes between the left and right subtrees of each node is calculated. If it exceeds a certain threshold, it often indicates a significant difference in the timing data of the invocations corresponding to the dimension used for classification, indicating the presence of an anomaly with high probability. The root cause locator triggered by anomalies ranks the root causes based on the frequency of occurrence of instances in the anomalous invocations.

**Deep Learning.** TraceAnomaly [22] encodes response time and invocation paths into a service trace vector and applies posterior flow to a variational autoencoder (VAE), enabling the model to capture the normal state of traces in a more accurate and robust manner. TraceAnomaly [22] performs anomaly detection tasks based on the reconstruction probability of VAE. With the detected anomalous trace paths, this method can narrow down the failure troubleshooting scope in an interpretable manner and determine the root cause. Similar to TraceAnomaly [22], TraceModel [76] also incorporates deep learning methods using VAE on top of ModelCoder [74]. Instead of using the traditional standard deviation band method for detecting anomalous traces like ModelCoder [74], TraceModel [76] trains a VAE for each request category using the response time of normal instances and calculates the average value and standard deviation of the reconstruction probabilities of these normal data. For the trace to be detected, when inputted into the corresponding VAE of its request category, if the reconstruction probability is below a threshold, it is determined as an anomalous trace. TraceModel [76] improves the accuracy of trace anomaly detection by mapping response time to reconstruction probabilities using VAE, and its root cause localization technique is consistent with ModelCoder [74].

*4.3.3 Techniques Based on Root Cause Analysis.* Root cause analysis is significant for completing failure diagnosis task that utilizes techniques from graph theory, probability statistics, causal analysis, and other related fields, building upon the foundation of anomaly detection. The effectiveness and accuracy of failure diagnosis heavily depend on the choice of root cause analysis techniques. There are simple yet effective methods such as techniques based on similarity matching [25, 74]. Researchers have also explored techniques based on spectrum used in the software testing domain to localize root causes [27, 56, 75, 105]. However, techniques based on spectrum only consider the normal and abnormal states of traces, overlooking the latency characteristics of microservices [55] or the differences in indicating root causes across different traces [21]. These limitations can be addressed by combining personalized PageRank methods [21, 55]. Some researchers have also attempted to address root cause localization from the perspective of causal analysis [54, 130].

**Similarity Matching.** FPDB [25] transforms system traces during failure periods into processing flow, which is a sequence composed of system events and records the sequence of component invocations during request processing. The transformed processing flow and corresponding failure information are stored in a failure profile database. To diagnose a failure, the similarity between the target failure and the stored processing flows is computed, and the KNN failure information is returned. This identified failure information is provided to operators as clues for manually inspecting the root cause. These similarity-based matching methods [25, 74] require a pre-built failure database. To cover a broader range of failures, FPDB [25] also designs a failure injection tool.

**Spectrum.** When given end-to-end traces, where a user request passes through multiple service instances in a microservices system, it is reasonable to apply spectrum-based techniques to localize the root cause. For example, T-Rank [75] detects trace anomalies based on the 3-sigma principle of latency data. It treats the normality of traces as the success or failure of test cases and calculates the suspicious scores of the microservice instances involved in those traces to provide a ranking of root causes. Another similar method TraceRCA [56] filters out effective metrics based on the anomaly severity of indicators before and after failures occur, discarding indicators with insignificant changes. Then, TraceRCA [56] calculates the average value and standard deviation of the effective metrics at the invocation level during the normal period. If the anomaly severity of an effective metric exceeds a threshold, the trace containing the invocation is considered abnormal. Following anomaly detection, TraceRCA [56] first identifies a set of suspicious microservices. This is because in practice, sometimes only traces involving specific sets of microservices are affected by failures, rather than individual microservices. TraceRCA [56] uses the Jaccard Index (JI) score [133], which considers the proportion of abnormal traces passing through a particular microservice set and the proportion of abnormal traces among all traces passing through that microservice set. Microservice sets with higher JI score are considered more suspicious. TraceRCA [56] also observes that if a microservice has both incoming and outgoing abnormal invocations, it is likely influenced by failure propagation. Therefore, to narrow down the scope of microservices for investigation, TraceRCA [56] also considers the in-set suspicious score of abnormal traces passing through each microservice set as one of the criteria for ranking root causes. TraceContrast [105] first extracts the critical path from the invocation chain, and then detects normal and abnormal paths based on the 3-sigma principle. The introduction of the eCSP algorithm [134] aims to mine contrast sequential patterns, which occur frequently in anomaly paths but not in normal paths. Finally, these patterns are ranked using SBFL. Minesweeper [27] does not explicitly use SBFL techniques, but the basic idea is similar, *i.e.*, root cause appears less frequently during normal periods and more frequently during abnormal periods. Minesweeper [27] deals with a contiguous sequence of events. It uses the PrefixSpan algorithm [135] to mine patterns from traces and then calculates the *precision* and *recall* of these patterns in the control group and test group. The root cause ranking of sequential patterns is based on the computed $F_1 - score$.

**Walk.** TraceRank [55] observes that when facing multiple microservice instances that appear in the same abnormal request and have similar service dependency relationships, using spectrum analysis alone may not accurately pinpoint the root cause, as they may have the same coverage information. Additionally, spectrum analysis only focuses on the normal and abnormal states of trace chains and overlooks the characteristics of the microservices themselves (*e.g.*, latency and status code). Therefore, TraceRank [55] introduces a personalized random walk method. First, HAC is used to aggregate traces with similar structures together. Then, K-means clustering is applied to divide each cluster of traces with similar structures into two classes based on their latencies. If the latency difference between the two clusters exceeds a certain threshold, it is considered as detecting an abnormal trace, triggering the root cause localization process. The root cause localization module then calculates a suspicious score of spectrum analysis for each service instance based on the normality, abnormality, and service dependency of the traces. By leveraging the similar patterns in latency between the frontend microservice and the root cause microservice, the forward, backward, and selfward transitions of the transition probability matrix are defined based on the similarity of latency between the frontend microservice and each microservice instance. This process derives the PageRank ranking for each service instance. Combining the rankings from both approaches, the final root cause ranking is determined. Unlike TraceRank [55], MicroRank [21] proposes a naive spectrum analysis method that treats each request at an equal level, ignoring the differences between different requests. MicroRank [21] divides traces into normal and abnormal categories by calculating an expected threshold for latency. MicroRank [21] believes that if an abnormal trace involves fewer service instances, the potential root cause should be narrowed down to a smaller scope and receive more attention. Due to the imbalance in the number of trace types, if an abnormal trace occurs less frequently, it should also receive more attention to prevent diversion from the trace types that occur frequently. Therefore, MicroRank [21] designs a trace coverage tree to represent the dependency relationship between requests and service instances and defines a transition matrix. A preference vector is defined based on the occurrence count of traces and the number of covered service instances. The PageRank results obtained from the trace coverage tree corresponding to normal and abnormal traces are used as the weights for the spectrum analysis, resulting in the root cause localization.

**Causal Analysis.** In typical cloud environments, operators often revert the versions or resource configurations of microservices to a known safe state while keeping other factors unchanged. If the system quality is restored, the changed microservice or resource configuration is considered a likely root cause. This causal analysis method is known as counterfactual queries. However, applying this method in a production environment can potentially incur performance or resource usage overhead. Therefore, Sage [54] utilizes the call relationship and latency data from historical traces to train a VGAE for generating the necessary counterfactuals for causal reasoning. Concurrently, it uses CBN to model the dependency relationships among metric nodes, latency nodes, and latent variables in the inter-service invocation process. In the root cause analysis phase, Sage [54] employs a two-level approach: it first uses service-level counterfactuals to localize the root cause. This involves sequentially restoring the metrics of each microservice to their normal states and using the VGAE based on the CBN structure to generate hypothetical end-to-end latency for two counterfactual scenarios. The microservice that, when restored to normal, leads to a significant improvement in the hypothetical end-to-end latency is identified as the root cause. Once a microservice node is identified, Sage [54] can continue with resource-level root cause localization by repeating the aforementioned counterfactual query process. However, Sage [54] relies on different graph topology to update the model. Therefore, Sleuth [130] chooses GNN to learn causal relationships in spans for root cause analysis, which aggregates messages from neighbors using permutation invariant functions. The follow up is consistent with Sage [54], where deep neural networks are used to model latency and other data, followed by executing counterfactual queries. Additionally, due to the model being independent of graph topology, a pre-trained Sleuth [130] model can be transferred to different microservice applications without any retraining or with few-shot fine-tuning.

*4.3.4   Summary.* Early researchers have demonstrated the value of analyzing runtime paths [123] and client request traces [99] in failure diagnosis tasks. With the widespread adoption of the concept of traces and the development of more systems for generating and collecting traces, some techniques [17, 74, 76] combine the instance deployment relationships in the topology with the service dependency graph extracted from traces to model the complex relationships between microservice nodes. Additionally, some techniques use business flow [1], processing flow [25], or events [27]. These are extracted or constructed from end-to-end traces, and essentially, a trace is a business flow, a processing flow, or a sequence of events. We summarize the failure diagnosis techniques through traces in Table 5 and classify them based on the emphasis on trace analysis processes in the papers. The table also provides information on the core methods and diagnostic targets.

Current studies on root cause localization [20, 21, 52, 55–58, 74–76, 105, 130] mainly focuses on service-level and instance-level. For the finer-grained component-level, some techniques different from logs and metrics focus on the operations or methods of a specific API request or invocation [1, 53, 129]. Other techniques focus on the path information reflected in traces [22, 25, 27]. TraceAnomaly finds the longest common path and considers the next called microservice of the longest common path as the possible root cause. FPDB [25] locates specific processing flows and code regions. A buggy code region can be a statement, a basic block, or an entire function, depending on the historical failure information stored in the failure profile database. Minesweeper identifies a trace as a sequence of events and diagnoses patterns of events that differ from the control group in the test group. Studies on failure classification [17, 52, 74, 76] mostly covers the failure types summarized in Section 3.3. It is worth noting that the latency recorded in traces can intuitively reflect latency-related failures, and a significant portion of other types also eventually manifest in the response time of microservices. Furthermore, the extracted call relationships from traces can intuitively reflect path-related failures and assist failure diagnosis in providing more interpretable results.

Analyzing the above techniques, we can glimpse the general steps of trace analysis, which include feature extraction, anomaly detection, and root cause analysis. In addition to the explicit response time, request status, request latency and specific information of different businesses recorded in traces, researchers can extract features related to upstream and downstream dependencies from the recorded call relationships [52, 53, 74]. Moreover, the degree of deviation from normality is also a common feature [57, 58, 129]. The known techniques for anomaly detection on traces can generally be divided into structural anomalies [1] and latency anomalies [20, 22, 76]. These techniques cover visual observation, machine learning, and deep learning aspects. However, there is still a long way to go from anomaly detection to failure recovery, and root cause analysis is an indispensable part of this process. Comparatively simple but effective techniques include similarity-based matching [25, 74] and spectrum analysis techniques [27, 56, 75, 105]. Additionally, some techniques, such as the personalized PageRank algorithm [21, 55], comprehensively consider the latency of microservices and the indicative capabilities for root causes. Apart from introducing spectrum analysis techniques from the software testing field, Sage [54] and Sleuth [130] attempts to address root cause localization from the perspective of counterfactual queries [54] and causal analysis.

Table 5. Summary of surveyed failure diagnosis techniques through traces, based on their *category*, the publication *year*, the *data* (*e.g.*, T for traces, E for events, and TP for topology) they use, the *principle* (*e.g.*, R for techniques based on rule, ML for techniques based on machine learning, and DL for techniques based on deep learning) and *supervision* (*e.g.*, Sup for supervised techniques, and Unsup for unsupervised techniques) type, the core *method*, and the diagnostic *target* (*e.g.*, S for service-level localization, I for instance-level localization, C for component-level localization, and FT for failure types).

| Category | | Technique | Year | Data | Principle | Supervision | Core Method | Target |
|---|---|---|---|---|---|---|---|---|
| **Feature Extraction** | Failures | MEPFL [17] | 2019 | T, TP | ML, DL | Sup | Random Forest + KNN + MLP | S, FT |
| | Upstream and Down-stream Dependencies | RanCorr [53] | 2009 | T, TP | R | Unsup | Weighted Power Mean | S, I, C |
| | | ModelCoder [74] | 2021 | T, TP | ML | Sup | 3-sigma + Particle Swarm + Optimization | I, FT |
| | | MicroHECL [52] | 2021 | T | ML | Sup | One Class SVM + Random Forest + 3-sigma | S, FT |
| | Anomalous Deviation | CloudDiag [129] | 2013 | T | ML | Unsup | Coefficient of Variation + RPCA | I, C |
| | | WinG [58] | 2022 | T | ML | Unsup | DTW | S |
| | | TraceNet [57] | 2023 | T | R | Unsup | 3-sigma | S |
| **Anomaly Detection** | Visualization | GMTA [1] | 2020 | T | R | - | Visualization | C |
| | Machine Learning | MicroSketch [20] | 2022 | T | ML | Unsup | Distributed Distribution Sketch + RRCF | I |
| | Deep Learning | TraceAnomaly [22] | 2020 | T | DL | Unsup | Deep Bayesian Networks with Posterior Flow | S C |
| | | TraceModel [76] | 2021 | T, TP | DL | Sup | VAE + 3-sigma + Particle Swarm Optimization | I, FT |
| **Root Cause Analysis** | Similarity Matching | FPDB [25] | 2016 | T, E | R | - | Edit Distance + Gaussian influence | C |
| | Spectrum | T-Rank [75] | 2021 | T | R | Unsup | 3-sigma + SBFL | I |
| | | TraceRCA [56] | 2021 | T | ML | Unsup | 3-sigma + FP-Growth + JI Score | S |
| | | TraceContrast [105] | 2024 | T | R | Unsup | Critical Path Extraction + 3-sigma + eCSP [134] + SBFL | C |
| | | Minesweeper [27] | 2021 | T, E | R | Unsup | PrefixSpan | C |
| | Walk | TraceRank [55] | 2023 | T | ML | Unsup | 3-sigma + HAC + K-means + SBFL + Personalized PageRank | S |
| | | MicroRank [21] | 2021 | T | ML | Unsup | 3-sigma + SBFL + Personalized PageRank | I |
| | Causal Analysis | Sage [54] | 2021 | T | DL | - | CBN + VGAE + Counterfactual Queries | S, C |
| | | Sleuth [130] | 2023 | T | DL | Unsup | BERT + JI Score + HDBSCAN + CBN + Counterfactual Queries | I |

In summary, traces can be regarded as a combination of graph and time series data. On the one hand, the call relationships recorded in traces naturally possess a graph structure. Therefore, many techniques focus on graphs and detect structurally anomalous call paths through visualization [1] or deep learning [22], which are failure types that are difficult to explicitly address with logs or metrics. Techniques based on spectrum [27, 56, 75] consider the coverage of edges on nodes in the graph. However, in traditional spectrum analysis techniques, traces have equal status, ignoring the differences in the indicative capabilities for root causes. To address this limitation, techniques based on walks [21, 55] use the call dependency graph to define a transfer matrix while incorporating spectrum analysis. Most of the above techniques focus on the situation of the node to be diagnosed, while some techniques [52, 53, 74] also consider the relevant features of upstream and downstream nodes in the call relationship. On the other hand, the various information recorded in traces, such as latency, can provide a perspective of time series analysis for failure diagnosis [17, 57, 58, 129]. MEPFL [17] defines a series of features directly related to failures, such as resource features (*e.g.*, memory and CPU consumption) that reflect the resource consumption of microservice instances and their nodes.

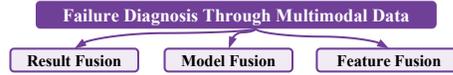

Fig. 7. Categories of failure diagnosis techniques through multimodal data.

### 4.4 Failure Diagnosis Through Multimodal Data

Currently, Solutions for failure diagnosis mostly rely on single-modal data. However, the limitations of techniques through single-modal data have been identified [19, 96]. They may result in ineffective diagnosis due to the inability to capture or entirely miss abnormal information caused by certain failures. To overcome the limitations of single-modal data, researchers have proposed numerous techniques through multimodal data to achieve more effective failure diagnosis, including logs, metrics, traces, events, and topology [16, 19, 26, 39, 50, 51, 73, 95, 96, 136]. Multimodal data combines information from various sources and provides a more comprehensive reflection of the system's operational state. It can identify issues across different aspects, adapt to and diagnose a wider range of failure types, and uncover more granular root causes. The integration of multimodal operational data for automated failure diagnosis has become a significant study focus in both academia and industry.

Our definition of multimodal data in the survey of failure diagnosis is as follows: Current failure diagnosis techniques primarily focus on the observability of microservice systems, with logs, metrics, and traces being the three pillars of observability [18]. Therefore, failure diagnosis through multimodal data should include at least two of these three types of data. As shown in Figure 7, we categorize them into result fusion, model fusion, and feature fusion.

*4.4.1 Result Fusion.* Result fusion techniques generally involve processing specific stage tasks, such as anomaly detection, based on single-modal data. Then, specific fusion analyses are performed between stage results of single-modal data [73] or between stage results and other modal data [95], considering various factors, to obtain the final diagnostic results. Result fusion techniques [73, 95] are commonly used in early stage.

ICWS'20 [95] considers the correlation between logs and metrics. Specifically, ICWS'20 [95] first uses the Deeplog [137] algorithm for log anomaly detection to obtain log anomaly scores. This stage result, which focuses on logs, reflects the system's level of anomalies. ICWS'20 [95] then utilizes mutual information (MI) to calculate the correlation between the log anomaly scores and metrics, aiming to locate the root cause components. PDiagnose [73] is the first work that utilizes logs, metrics, and traces for failure diagnosis. PDiagnose [73] starts by analyzing the metrics and detecting abnormal metrics using KDE and weighted moving average (WMA). And then, PDiagnose [73] creates an anomaly queue and extracts partial features from it. The comparison results between these features and thresholds serve as the primary basis for determining if the system has anomalies. For the traces, PDiagnose [73] uses the messages structured as *<Req, Caller, Callee, Duration>* to represent the calling relationships between microservices and the execution time of requests. PDiagnose [73] reports suspicious microservices by comparing the duration with specific thresholds, considering it as one of the criteria for determining if the system has anomalies. After obtaining the root cause service through a voting mechanism, PDiagnose [73] categorizes the problem into subsystems using the anomaly queue and determines the anomalous subsystem based on the proportion of abnormal metrics. Finally, PDiagnose [73] performs a double-check by examining the metrics and analyzing log entries within the subsystem that contain keywords such as *error* and *problem* to identify the root cause components and report suspicious logs. These findings are added to the root cause indicators as the final diagnostic result.

*4.4.2 Model Fusion.* Model fusion techniques extract different feature information from multimodal data, which serves as input for specific models. This requires models to possess the ability to handle and learn from diverse features [16, 26, 50].

Graph theory models exhibit the capability to effectively integrate multimodal data. Both Groot [26] and TrinityRCL [50] adopt similar approaches. Groot [26] constructs the service dependency graph using traces and logs. It then fuses performance metrics, status logs, and developer activities to generate events. Events and causal rules are integrated into the service dependency graph to establish a causality graph. When an alert is triggered, the causality graph is utilized as input to the GrootRank algorithm, which is a personalized PageRank algorithm, to obtain a ranked list of root causes for the events. TrinityRCL [50] is a root cause localization method utilizing multimodal data for multi-granularity localization. It defines four levels of root cause localization granularity, namely application-level, service-level, host-level, and metric&code-level. TrinityRCL [50] collects data via the APM system Raptor, performs anomaly detection, and transforms log entry counts into temporal data. It recursively searches for affected services from abnormal ones to get dynamic call relationships and calculate call failure rates. Based on these relationships, it incorporates various nodes to construct a dynamic causality graph with temporal data. Finally, TrinityRCL [50] localizes root causes at different levels from this graph using RWR.

MicroCBR [16] proposes a failure diagnosis framework that combines offline updates with online diagnosis based on case-based reasoning, which integrates multimodal data into a knowledge graph. In the offline phase, MicroCBR [16] constructs a spatio-temporal fault knowledge graph by embedding the existing knowledge base into a topology graph of system instances. During the online diagnosis phase, anomalies detected from multimodal data are converted into fault fingerprints and embedded into the graph. Subsequently, by allocating anomaly weights and hierarchical case-based reasoning with historical data, failure reports are generated to help operators identify failure types and update the knowledge base. UniDiag [138] introduces a failure diagnosis framework for microservice systems that leverages temporal knowledge graphs (TKGs) to integrate multimodal data. It constructs TKGs to capture system dynamics and uses microservice-oriented graph embedding (MOGE) to model structural and temporal relationships. In the offline phase, UniDiag [138] clusters graph embeddings to identify failure patterns, reducing annotation effort by labeling only cluster centers. During online diagnosis, anomalies are matched to clusters or used to create new ones.

*4.4.3 Feature Fusion.* Feature fusion techniques first process multimodal data to extract a unified feature matrix [39, 51, 136] or unified event representation[19, 96] as input for the models. The feature fusion methods only need to handle a specific unified representation, and the model's output results serve as the final diagnostic outcome.

CloudRCA [39] utilizes time series anomaly detection and log clustering modules to process metric and log data. Specifically, CloudRCA [39] designs a simplified RobustPeriod [139] algorithm to determine the periodicity of time series. Then, the RobustSTL [139] algorithm is applied to decompose the time series, and anomaly detection is carried out on the decomposed series using a divide-and-conquer approach to obtain anomaly sequences. For log processing, CloudRCA [39] designs an adaptive frequent template tree (AFT-tree) based on FT-tree [140] to extract log templates, and Word2vec is employed for vectorization. Based on cosine similarity between vectors, CloudRCA [39] applies HAC to obtain log patterns. The anomaly metric sequences and log pattern sequences are then integrated to form a unified feature matrix. Finally, CloudRCA [39] combines the feature matrix with module dependency relationships to construct a knowledge-informed hierarchical Bayesian network (KHBN) for online root cause analysis.

Eadro [51] integrates temporal processing and feature extraction of logs, metrics, and traces. Logs are parsed into template events using Drain [141], and their temporal intensity is modeled via Tick [142] with a Hawkes model. Metrics and trace-derived latencies are processed through a one-dimensional dilated causal convolution (1D DCC) [143] to capture temporal and cross-series dependencies. These features are fused nonlinearly with gated linear units (GLU) [144] and combined with a trace-derived dependency graph as input to graph attention networks (GAT) for state representation learning. Eadro [51] highlights the link between anomaly detection and root cause localization, enabling the model to assess system normalcy or rank potential root causes. Similarly, in the modality fusion stage, Medicine [145] utilizes a multimodal fusion module with channel attention to combine the raw statistical features of different modalities and the failure classification features extracted by feature encoders. What sets Medicine [145] apart is its design of modality-specific feature encoders tailored to the unique characteristics of metrics, logs, and traces, to capture complementary information between modalities. Medicine [145] employs a modality evaluation component to distinguish between high and low-yield modalities. High-yield modalities are optimized and suppressed through a gradient suppression component, while low-yield modalities are enhanced through a feature enhancement component. These components work together to dynamically adjust the training process, ensuring that each modality contributes effectively to the overall model performance. MULAN [136] performs unified temporal processing on logs and metrics. To prevent the potential loss of valuable insights caused by extracting invariant information separately, MULAN [136] employs GraphSage [146] to extract both individual specific representation and a joint invariant representation. These representations are then fused using a contrastive learning approach to obtain a fused log and metric representation. Additionally, MULAN [136] constructs learnable causal graphs for each representation. Subsequently, MULAN [136] introduces a KPI-aware attention module to incorporate the decoder results for causal graph fusion. Finally, MULAN [136] utilizes RWR to localize the root cause.

DiagFusion [19] and Nezha [96] both transform multimodal data into unified event representations, but they adopt distinct approaches in event design and extraction. DiagFusion [19] uses historical failures to train an event extraction model, performing anomaly detection on metrics and traces, and log parsing on logs to generate tuple-formatted events. These events are sequenced chronologically to achieve data fusion. GNN is then trained using the event sequence and a dependency graph derived from traces and topology. During real-time diagnosis, the trained event embedding model and GNN are employed to identify root cause instances and failure types. Nezha [96] differs from DiagFusion [19] in the temporal processing approach for event extraction, better preserving the execution context and enhancing the interpretability of root cause analysis. For logs, it integrates internal and inter-service information by inserting a *traceID* and parsing log messages to extract log events. For metrics, it detects anomalous alarms to generate alarm events. For traces, it represents event messages by concatenating span names with *start*, *end*, and *asyn* strings, considering both synchronous and asynchronous call relationships. After unifying multimodal data into event representations, Nezha [96] employs an expected pattern ranker and an actual pattern ranker to explore event patterns. The expected pattern ranker identifies patterns during fault-free phases to determine failure root causes, while the actual pattern ranker identifies deviations during fault-suffering phases. Finally, a pattern aggregator filters redundant patterns to finalize the root cause ranking.

DeepHunt [147] and ART [148] both unify multimodal data into time-series formats and leverage self-supervised learning to train models that compute reconstruction errors for downstream failure diagnosis tasks. Specifically, DeepHunt [147] employs a GAE for reconstruction tasks and combines temporal and call dependencies to establish a root cause score, with a feedback mechanism enabling continuous model optimization. ART [148], on the other hand, utilizes Transformer encoder, gated recurrent unit (GRU), and GraphSAGE [146] to model channel, temporal, and call dependencies, predicting system snapshots at the next time step. The resulting unified and semantically rich failure representations enable multiple downstream tasks.

Table 6. Summary of surveyed failure diagnosis techniques through multimodal data, based on their *category*, the publication *year*, the *data* (*e.g.*, L for logs, M for metrics, T for traces, E for events, and TP for topology) they use, the *principle* (*e.g.*, R for techniques based on rule, ML for techniques based on machine learning, and DL for techniques based on deep learning) and *supervision* (*e.g.*, Sup for supervised techniques, and Unsup for unsupervised techniques) type, the core *method*, and the diagnostic *target* (*e.g.*, S for service-level localization, I for instance-level localization, C for component-level localization, and FT for failure types).

| Category | Technique | Year | Data | Principle | Supervision | Core Method | Target |
|---|---|---|---|---|---|---|---|
| Result Fusion | ICWS'20 [95] | 2020 | L, M | ML | Unsup | Oversampling Anomalies + Adding Gaussian Noise + MI | C |
| | PDiagnose [73] | 2021 | L, M, T | ML | Unsup | Anomaly Queue + Voting Mechanism | I, C |
| Model Fusion | Groot [26] | 2021 | L, M, T, E | R, ML | Unsup | Event Causal Graph + PageRank | C |
| | TrinityRCL [50] | 2023 | L, M, T | R | Unsup | Causality Graph + RWR | S, I, C |
| | MicroCBR [16] | 2022 | L, M, T, TP | R | Unsup | Knowledge Graph + Hierarchical Case-based Reasoning | FT |
| | UniDiag [138] | 2024 | L, M, T, TP | ML, DL | Unsup | TKG + MOGE + HAC | FT |
| Feature Fusion | CloudRCA [39] | 2021 | L, M, TP | ML, DL | Sup | Unified Feature Matrix + KHBN | I, FT |
| | Eadro [51] | 2023 | L, M, T | ML, DL | Sup | Multi-modal Fused Representation + GAT | S |
| | Medicine [145] | 2024 | L, M, T | ML, DL | Sup | Modality-specific Feature Encoders + Multimodal Adaptive Optimization | FT |
| | MULAN [136] | 2024 | L, M | DL | UnSup | Contrastive Learning + GraphSage [146] + KPI-Aware Attention + RWR | I |
| | DiagFusion [19] | 2023 | L, M, T, TP | DL | Sup | Unified Event + GNN | I, FT |
| | Nezha [96] | 2023 | L, M, T | ML | Unsup | Unified Event + Event Pattern Mining and Comparison | C |
| | DeepHunt [147] | 2024 | L, M, T, TP | DL | Unsup | Self-supervised Learning + GAE + MLP | I |
| | ART [148] | 2024 | L, M, T, TP | DL | Unsup | Self-supervised Learning + Transformer Encoder + GRU + GraphSAGE | I, FT |

*4.4.4 Summary.* To adapt failure diagnosis techniques to a wider range of failures and achieve more granular root causes, the integration of multimodal data for automated failure diagnosis has become an important direction focus in both academia and industry. Using multimodal data allows for a comprehensive representation of system states and captures failure patterns, leading to improved diagnostic effectiveness and interpretability [19, 96]. Table 6 summarizes the fault diagnosis methods based on multimodal data that we survey and categorize them based on the proposed fusion strategy. In addition, Table 6 provides the required modalities, core methods, and targets of the surveyed techniques.

Regarding the techniques on root cause localization [19, 26, 39, 50, 51, 73, 95, 96, 136, 147, 148], different techniques focus on different target systems at various levels of granularity. Some techniques can even localize root causes at different levels of granularity [50, 73, 136], where the component-level includes metrics, log messages, and unified event representation [19, 96]. As for failure classification [16, 19, 39, 138, 145, 148], it essentially covers the failure types compiled in Section 3.3.

Compared to techniques based on single-modal data, failure diagnosis based on multimodal data starts relatively late, and its main challenge lies in effectively integrating heterogeneous multimodal data. Result fusion techniques have not made significant improvements compared to single-modal approaches and are generally used in earlier studies. Result fusion techniques typically start by performing anomaly detection or similar operations on individual modal data. Then they localize root causes based on the correlation between anomaly scores and metrics [95] or through a voting mechanism that balances the results of multimodal processing [73]. PDiagnose [73] is the earliest publication to use three modalities of logs, metrics, and traces for failure diagnosis. Subsequent papers [19, 51, 96, 136] on extracting unified representations of multimodal data was inspired by this work. Given the need for models to handle various features, model fusion techniques often require more complex designs. Graph-based models, which excel at combining multimodal data, are widely used. They typically construct a causality graph [26, 50] or knowledge graph [16, 138] based on the topology or call information in traces. The graph's nodes generally store feature information from multimodal data and diagnostic results are obtained through graph traversal [26, 50] or inference [16].

Recent studies often convert multimodal data into a unified feature representation. Initially, specific preprocessing operations are applied to heterogeneous multimodal data to extract essential information and obtain feature matrix [39, 51, 136, 145, 147, 148] or unified fusion events [19, 96]. Methods for feature fusion often aim to achieve a more reasonable representation of unified features, obtaining more important multimodal feature information while maintaining better interpretability. These unified feature representations are then used as inputs to deep learning or other machine learning methods for failure diagnosis.

## 5   DISCUSSION

The surveyed failure diagnosis techniques can localize root causes at different levels or determine failure types on microservice systems. Based on the data used by these techniques, we provide detailed classification and summary. Next, we analyze the trends of failure diagnosis (Section 5.1). Then, we summarize from the literature and qualitatively discuss these technologies from various perspectives, specifically addressing practical requirements and outstanding challenges related to the granularity and explainability of failure diagnosis (Section 5.2), their characteristics and portability (Section 5.3), and the evaluation of accuracy and costs (Section 5.4). Lastly, we discuss the best current practices and future directions (Section 5.5).

### 5.1   Trends and Adavancements

Early failure diagnosis techniques [1, 53, 80, 99, 109, 123] extract or focus on relatively simple features for correlation or visualization analysis. However, as the complexity of modern microservice systems continues to increase, these techniques have shown a significant decline in diagnostic effectiveness due to the overwhelming amount of data. Moreover, they often rely heavily on extensive expert experience and manual efforts, making them susceptible to subjective factors and resulting in inconsistent diagnostic outcomes. The development of AI technology has brought new opportunities for failure diagnosis. Recent techniques [21, 22, 55, 59, 60, 76, 107, 108] aim to achieve automated and intelligent failure diagnosis by leveraging machine learning and deep learning methods, minimizing or eliminating the reliance on human factors.

Traditional failure diagnosis techniques typically focus on single-modal data. However, multimodal data, including logs, metrics, traces, events, and topology, provides a comprehensive understanding of the system's state, enabling failure diagnosis techniques to yield more precise results. The limitations of single-modal failure diagnosis techniques have become increasingly evident, leading to a growing body of research on how to correlate and integrate multimodal data and extract key information for failure diagnosis. In Section 4.4, we categorize multimodal failure diagnosis techniques into result fusion, model fusion, and feature fusion. Result fusion techniques [73, 95], proposed earlier, have relatively lower requirements for accuracy and universality. While these techniques are simple to implement, they may yield different diagnostic results when modeling individual data, making it challenging to make decisions for the final diagnosis. Additionally, maintaining separate models for each data incurs significant costs in terms of storage and maintenance. Therefore, these techniques do not possess clear advantages over single-modal failure diagnosis techniques and do not address the limitations of single-modal data. Model fusion techniques [16, 26, 50], on the other hand, strive for higher accuracy and better universality by incorporating more observable data to achieve a comprehensive representation. However, they are constrained by the heterogeneity of multimodal data. Advanced research has proposed methods to unify the representation of heterogeneous data, laying the foundation for feature fusion. Feature fusion techniques [19, 39, 51, 96] enhance failure diagnosis by integrating features from multiple modalities, aiming for a more unified representation of heterogeneous data and demanding improved interpretability in failure diagnosis.

## 5.2 Granularity and Explainability

In Section 3.3, we summarized the granularity of failure diagnosis information provided by the investigated diagnostic techniques. Coarse-grained diagnostics may only indicate overall issues with a service, which is suitable for simple systems or initial rapid assessments. However, they fail to pinpoint the specific instances or components where the failures occur, making troubleshooting challenging. On the other hand, finer-grained diagnostics typically involve more complex data processing and analysis but can help engineers resolve and repair failures more accurately and quickly.

Generally, the diagnostic granularity of a technique is determined during its model design phase. However, there are also techniques with relatively flexible diagnostic granularity [50, 71]. They share a common approach of constructing a causal relationship graph and then searching for root cause nodes in the graph. The granularity of root cause localization depends on the types of nodes set during graph construction, which can be services, instances, metrics, codes, or other components. At the component-level, metrics indirectly reflect the state through variable values, while logs and traces explicitly and directly record the information. For one thing, developers often embed log statements within programs to print system runtime states, errors, information, and more. Therefore, logs often serve as a reflection of the system's behavior, especially the identified log lines [78, 79, 84, 85, 88] and events [82, 86, 87]. For another thing, traces not only record the interaction between services or instances but also include the operations or methods specific to an API or request [1, 53, 129], providing data support for finer-grained failure diagnosis.

The explainability in failure diagnosis refers to the extent to which humans can understand the causal relationship between the input features of the model and the diagnostic outputs. This is crucial for understanding the root causes, formulating repair strategies, and taking appropriate actions. An interpretable model enables stakeholders, including experts and non-experts, to understand the logic behind the model's decisions. It requires the model to explain why certain information or features are more relevant to root causes and why they are diagnosed as the most probable root causes.

Comparison is a common solution to ensure model interpretability by providing historical states as references for diagnostic results. Specifically, some techniques rely on a core component of interpretability, which is the historical failure repository [25, 59, 74, 76, 78, 79, 81, 92, 104, 107, 108]. By comparing the similarity between system failures, relevant historical failures along with their corresponding root causes and mitigation steps can be obtained. Referencing historical root cases often leads to more trustworthy diagnostic results. it is worth noting that comparing the differences between the system and the expected normal state can also provide intuitive and interpretable results. Additionally, causality analysis can naturally introduce interpretability into failure diagnosis, including causal reasoning [94] and counterfactual queries [54, 130]. Compared to black-box models, step-by-step inference from phenomena to source can better demonstrate the logical reasoning process and interpretability of diagnostic results.

## 5.3 Characteristic and Portability

Logs record various events and operations that occur during system runtime, providing detailed contextual information. Metrics offer quantifiable measurements of key performance aspects of the system. By monitoring metrics, operators can assess the system's health and identify performance bottlenecks in real time. Traces capture the propagation process of requests within the system and the dependencies between different instances, which can be used to pinpoint the failures and their impact, enabling quick tracing of failure propagation paths. Based on these characteristics, logs and traces can help understand the operations, and propagation paths, providing relatively interpretable results. However, they lack a high-level overview of the system and require further investigation and analysis to address and repair failures. Moreover, the volume of logs and traces is often much larger than metrics, requiring significant time and computational resources for effective data management and processing. The diversity in formats and content of logs and traces also presents challenges. On the other hand, metrics offer real-time detection of potential failures due to their intuitiveness and quantifiability. However, they lack contextual information, resulting in poorer interpretability regarding the impact of failure propagation. Therefore, improving the comprehensiveness, accuracy, and interpretability of failure diagnosis is achieved by integrating information from different data, and comprehensively analyzing the system's state and behavior. Failure diagnosis through multimodal data requires more complex data integration and analysis techniques, as well as increased computational resource support.

Failure diagnosis solutions with good portability possess multiple advantages. Firstly, such solutions should have system and platform compatibility, enabling them to run on different operating systems and hardware architectures, simplifying the deployment and maintenance processes. Secondly, they can be configured and customized according to specific requirements to achieve optimal performance and effectiveness. This means operators can tailor the solutions to better adapt to their specific environments and requirements. Furthermore, the scalability and modular design of these solutions are also crucial for achieving portability. Lastly, due to the uniqueness of fault diagnosis tasks, microservice systems monitor various data in real time, and each data has diverse formats and content. The solutions should have general data processing capabilities and analytical mechanisms to ensure adaptation to different data formats and structures. One of the most typical designs for portability is the logical graph module. Techniques [16, 39, 50, 71, 130] build graphs based on logical consensus rather than specific calling dependency or deployment relationships. Therefore, analysis processes or techniques based on logical graphs can often switch between different scenarios at a lower cost.

## 5.4 Accuracy and Costs

Higher accuracy is crucial for reducing failure recovery time and improving system availability. Additionally, it can prevent the erroneous replacement or repair of components that are not faulty. However, false positives and false negatives are common issues in practical scenarios. A high false positive rate leads to the operations team wasting time and resources dealing with false alerts, reducing attention to real failures. On the other hand, a high false negative rate results in missed opportunities for diagnosing and resolving real failures, prolonging failure recovery time and increasing system downtime. Reducing the false positive rate may increase the false negative rate, and vice versa. Therefore, researchers are attempting to optimize them. Failure diagnosis through multimodal data combines complementary information from different data sources, effectively improving accuracy and reliability. Furthermore, some techniques [36, 37, 42, 92, 94, 98] attempt to actively involve human experts in the training and improvement process of models. Particularly, failure diagnosis often faces complex scenarios and boundary conditions, and human-in-the-loop training establishes a feedback loop that allows human experts to continuously review and correct the model's output. These techniques have significant significance and potential to drive the development of future failure diagnosis.

Unknown failure diagnosis is another important issue. For unknown failures, existing models may not be able to make accurate judgments or provide reliable results because they are trained and learned based on known failure cases. Some techniques have explored different solutions. For example, Log3C [83] expands the diagnostic target from specific known failure types to impactful service system problems. By analyzing the correlation between different clusters of log sequences and user-perceived metrics, it recommends the most likely clusters of log sequences that lead to system performance degradation. MicroCBR [16] focuses on generating detailed failure reports, combining results from multimodal data with historical failure information. failure diagnosis essentially maps the results to root causes and failure types. When designing, researchers keep a continuous focus on this mapping layer while also attempting to obtain finer-grained results in the pre-mapping stage. This provides more details in the face of unknown failures or when the mapping layer based on known failures becomes ineffective.

When selecting or deploying appropriate techniques, it is necessary to consider both time and space costs and make trade-offs based on the requirements of the practical scenario. In terms of time cost, the efficiency should meet the requirements of real-time or near real-time diagnosis. These techniques should use a shorter history data to quickly obtain usable models. In terms of space cost, the storage cost generated by the size and number of model parameters should be compatible with the available resources, and the scheduling cost during deployment and maintenance should be feasible. To address this requirement, some techniques [52, 65] tend to choose simple data analysis or methods over complex deep learning models when they achieve similar results. Other techniques [70, 83] attempt to consider the costs in the algorithm design process.

Although these techniques can automate and intelligently assist the failure diagnosis process, human involvement is still necessary. The automation level has been continuously improving in terms of result interpretation [25, 54, 59, 74, 76, 78, 79, 81, 92, 94, 96, 104, 107, 108, 130] and model update maintenance [16, 39, 50, 71, 130]. However, in terms of failure labeling and model training optimization, the cost of human labor is an unavoidable and important consideration. Supervised techniques require a sufficient number of labeled data to train the model, and the quality of labeling is crucial to the performance. Labeling operators need to carefully inspect each data sample and assign appropriate failure labels to them. This is a time-consuming and tedious process. Additionally, incorporating human feedback during model iteration and optimization also incurs some costs. To balance model performance and human costs, on the one hand, we need to achieve the highest quality of feedback with minimal human effort, and on the other hand, using large language models to obtain feedback is also a good option.

## 5.5 Best Current Practices and Future Directions

Based on the above discussion, we offer insights into probable best practices for different scenario requirements to assist on-call engineers in selecting the most suitable technique. (1) In terms of diagnostic granularity and explainability, DéjàVu[92] and Nezha [96] use decision trees or simple pattern comparisons to interpret the outputs of deep learning models, making the diagnostic results more transparent. These techniques are particularly suitable for environments that require traceable decision support. (2) To improve the adaptability and portability of failure diagnosis, a series of logic graph-based techniques [16, 39, 50, 71, 130] build graphs from widely accepted logical consensus, allowing failure diagnosis solutions to seamlessly switch across different platforms and environments. These techniques are all excellent options. (3) To balance high accuracy with cost control, $\epsilon$-diagnosis [65] uses a two-sample testing algorithm to simplify metric analysis, successfully finding a balance between accuracy and computational overhead. It avoids the complexity of training models. However, if data analysis is ineffective, machine learning or deep learning models should still be employed for more effective modeling.

Future research directions need to make breakthroughs and innovations in several key areas. (1) Currently, the results of failure diagnosis often require professional engineers to interpret, which may make it difficult for non-experts to effectively participate in fault handling and analysis. Therefore, improving the interpretability and user-friendliness of models, so that non-professional personnel can understand diagnostic results and make informed decisions, will be a crucial task. (2) The challenges related to data acquisition quality and granularity of data collection remain significant in the deployment of current diagnostic systems. Future research should focus on how to maintain efficient fault diagnosis capabilities despite unstable data quality, noise, or missing data, especially in systems with coarser data granularity. (3) Although current failure diagnosis techniques have become more automated, the feedback mechanism of "human in the loop" remains a key approach to improving model performance. Future research should further explore how to optimize human feedback to enhance model accuracy and reliability, reducing human intervention while ensuring that the system can continuously learn and improve. In-depth research in these directions will help advance intelligent fault diagnosis technologies toward greater efficiency and reliability, meeting the needs of increasingly complex microservice systems.

Recent years have seen significant progress in large language models (LLMs) in academia and industry, with models like ChatGPT gaining widespread attention. State-of-the-art LLMs have proven effective in addressing failure diagnosis problems [149–151]. A key focus is leveraging technologies such as knowledge graphs and LLMs for failure diagnosis. Knowledge graphs can establish structured relationships between system architecture and network topology, revealing underlying connections. LLMs excel in natural language processing tasks like relational reasoning, knowledge extraction, and similarity-based corpus queries [152]. However, LLMs alone struggle to provide topological explanations for real-time monitoring data. Integrating statistical analysis findings into knowledge graphs and using this accumulated knowledge to enhance LLM learning efficiency creates a cycle of mutual development. This synergy enables both step-by-step and end-to-end failure diagnosis approaches.

## 6 DATASETS AND METRICS

### 6.1 Datasets and Toolkits

High-quality large-scale datasets provide important experimental scenarios and evaluation standards for algorithm innovation and technological advancement, promoting the integration and innovation of knowledge across disciplines. Similar to other data-driven domains, datasets play a crucial role in the domain of failure diagnosis in microservice systems. Industrial microservice systems are characterized by complex service relationships and massive underlying resources. In contrast, the academic community often lacks access to real-world data and market-driven factors, resulting in incomplete or unobtainable content and a lack of practical applications in industry. This poses challenges for empirical studies in microservice system failure diagnosis, as they often lack high-quality datasets and appropriate toolkits to support automated or semi-automated validation and evaluation. To strengthen research in failure diagnosis, it is necessary for the industry to provide production-level multimodal data (*i.e.*, logs, metrics, traces, events, and topology), for the academic community to participate in annotating rich datasets, and for advanced failure diagnosis solutions to be publicly available. Collaborative efforts are needed to advance the construction of high-quality standard datasets and provide unified evaluation criteria. After carefully searching, we have organized a list of publicly available failure diagnosis datasets and toolkits, as shown in Table 7 and 8.

Currently, several publicly available datasets have been used by researchers to validate the effectiveness of their failure diagnosis techniques. These datasets not only facilitate the reproducibility of experimental results in papers but also provide valuable data resources for future studies. The AIOps Challenge datasets [21, 55, 57, 58, 73, 74, 76, 147, 148] are collected by a research team from Tsinghua University from production environments such as large wireless service providers, commercial banks, and e-commerce systems. The international AIOps challenge competitions, which have been held since 2018, also encourage researchers in related fields to conduct more in-depth research, promoting the development of the relevant domains. The Generic AIOps Atlas dataset (GAIA) [19, 138, 145] mainly comes from the MicroSS microservice simulation system developed by Cloud Wisdom. It contains authorized and rigorously anonymized user data from Cloud Wisdom. Additionally, it simulates failures that may occur in real systems and provides records of all failure injections. It consists of over 6,500 metrics spanning two weeks, 7 million log entries, and detailed traces. Some papers [51, 92, 96] use multimodal data collected from the publicly available platform TrainTicket for experimentation and mimic application failures using failure injection tools.

Table 7. Summary of publicly available datasets.

| Name | Data | Details |
|---|---|---|
| AIOps Challenge 2020 | M, T | The dataset collected by China Mobile Zhejiang from a real-world production microservice system. |
| AIOps Challenge 2021 | L, M, T, TP | The dataset collected by Tsinghua University from two large commercial banking systems. |
| AIOps Challenge 2022 | L, M, T, TP | The dataset collected by Tsinghua University from Hipster Shop. |
| GAIA | L, M, T, E, TP | Network performance, system metrics, application logs, and user activity of the MicroSS microservice simulation system developed by Cloud Wisdom. |
| TrainTicket-DéjàVu | M, TP | Metrics, failures and topology collected on the TrainTicket testbed. |
| TrainTicket-Eadro | L, M, T | Application logs, metrics and traces collected on the TrainTicket testbed. |
| TrainTicket-Nezha | L, M, T | Application logs, metrics and traces collected on the TrainTicket testbed. |
| Loghub | L | A large collection of log datasets from various systems. |
| SWaT, WADI | M | A series of sensor feature metrics collected by Singapore University of Technology and Design from a water treatment and distribution testbed. |
| SMAP, MSL | M | Feature metrics of different entities from spacecraft telemetry data. |

There are also some datasets in non-microservice systems that are worth mentioning. The Loghub dataset is compiled by a research team from the Chinese University of Hong Kong, aggregating log datasets from various types of systems in both real production environments and laboratory simulation environments. The maintained log dataset has a total size of over 77 GB and includes distributed system logs, supercomputer cluster logs, operating system logs, mobile application logs, server application logs, and standalone software logs. For metrics, the Singapore University of Technology and Design collected the secure water treatment (SWaT) dataset and water distribution (WADI) dataset from a water treatment testbed and a water distribution testbed, respectively. Similarly, NASA collected the soil moisture active passive satellite (SMAP) dataset and Mars Science Laboratory rover (MSL) dataset from spacecraft telemetry data, which have annotated records of anomaly sequences and types. These papers we surveyed also used the aforementioned non-microservice system log and metric datasets [32, 33], which can be utilized to validate the effectiveness and impact of failure diagnosis solutions.

In addition to the datasets, some publicly available testing platforms have made significant contributions to the development of related research. OpenStack [59, 78, 81] is a cloud computing platform project aimed at providing IaaS solutions. Hadoop [60, 88, 104, 108, 153] is a distributed computing framework for storing and processing large-scale data. Hadoop-based Spark [80, 84] refers to Spark applications running on a Hadoop cluster and represents an improvement over traditional MapReduce. RUBBoS [88] is a bulletin board benchmark that models an online news forum. Researchers can deploy HTTP servers, application servers, and a database server on this benchmark. They can then simulate several thousands of concurrent requests to different services and obtain logs from multiple services to form a dataset. The IBM Cloud Testbed [46, 47, 61, 78, 88] is an open cloud environment for experimentation and testing. This testing environment is based on the IBM Cloud platform and provides a variety of cloud resources and services such as virtual machines, containers, storage, databases, and artificial intelligence. It allows developers to easily create and manage their testing environments. Hipster-shop [20, 21, 93] is a cloud-native microservice e-commerce platform developed using technologies such as Kubernetes, Istio, Prometheus, and Grafana, showcasing microservice architecture patterns. It encompasses microservices for product catalog, shopping cart, payment, order processing, and front-end/back-end interfaces. Sock Shop [17, 35, 69], a simulated online sock shopping microservice application, demonstrates cloud-native development and deployment. It includes microservices like product catalogs, shopping carts, inventory, and payment services, each focusing on specific business functions and communicating via REST APIs. TrainTicket [22, 51, 92, 96, 100, 105, 136] is a benchmark system for train ticketing developed by a research team from Fudan University. Users can check, book, and pay for train tickets using TrainTicket [51, 92, 96, 100, 105, 136]. It consists of 24 microservices that actively interact with each other, similar to real-world industrial microservice systems.

Table 8. Summary of publicly available failure diagnosis toolkits.

| Name | Year | Data | Details |
|------|------|------|---------|
| DISTALYZER [86] | 2012 | L | An automated tool to investigate the performance issues in distributed systems. |
| FDiagV3 [87] | 2015 | L | An extended version of the FDiagV3 diagnostics toolkit for log files. |
| Log3C [83] | 2018 | L | A clustering-based technique to promptly and precisely identify impactful system problems. |
| SwissLog [60] | 2022 | L | A robust anomaly detection and localization tool for interleaved unstructured logs. |
| LogKG [107] | 2023 | L | A novel framework based on knowledge graphs. |
| DBSherlock [97] | 2016 | M | A heuristic failure classification technique optimized by historical metrics and user feedback. |
| Sieve [40] | 2017 | M, TP | A platform to derive actionable insights from monitored metrics in distributed systems. |
| DyCause [70] | 2021 | M | A causal failure diagnosis technique via sliding windows and crowdsourced graph fusion.. |
| DéjàVu[92] | 2022 | M, TP | An interpretable, actionable technique for localizing recurring failures in online service systems. |
| CIRCA [42] | 2022 | M, TP | A novel unsupervised causal inference-based failure localization technique. |
| RootCLAM [44] | 2023 | M, TP | A technique for causal-based root cause localization and anomaly mitigation. |
| TraceAnomaly [22] | 2020 | T | Unsupervised anomaly detection via novel trace representation and deep Bayesian networks. |
| TraceRCA [56] | 2021 | T | A practical root-cause microservice localization technique via trace analysis. |
| MicroCBR [16] | 2022 | L, M, T, TP | A framework with offline spatio-temporal graph construction for online troubleshooting. |
| UniDiag [138] | 2024 | L, M, T, TP | A failure diagnosis technique, leveraging TKGs to fuse multimodal data. |
| Eadro [51] | 2023 | L, M, T | A framework for troubleshooting via anomaly detection and root cause localization. |
| Medicine [145] | 2024 | L, M, T | A modal-independent framework based on multimodal adaptive optimization. |
| DiagFusion [19] | 2023 | L, M, T, TP | An automatic failure diagnosis technique extracting and unifying events from multimodal data. |
| Nezha [96] | 2023 | L, M, T | An interpretable and fine-grained RCA technique based on multimodal data. |
| DeepHunt [147] | 2024 | L, M, T, TP | An interpretable method with reconstruction errors, dependency modeling, and user feedback. |
| ART [148] | 2024 | L, M, T, TP | A unified framework using multimodal dependency modeling and failure representations. |

Failure diagnosis toolkits can save users the time and cost of reproducing experiments, thereby promoting the development of related research. Additionally, when dealing with highly complex microservice systems, publicly available toolkits make it easier and more widespread to conduct experiments in real-world application scenarios. The feedback obtained from experiments conducted on different systems helps researchers better determine the next steps for improvement and research directions.

Up until now, there was a scarcity of research on publicly available toolkits, accounting for only 21.43% of the total. This hindered the reproducibility and improvement of subsequent research work. However, in recent years, the number of studies on publicly available toolkits has increased significantly. In particular, Eadro [51] and Nezha [96] have open-sourced both their toolkits and the datasets used, which undoubtedly contributes to the advancement of failure diagnosis studies. In conclusion, Table 9 provides a compilation of the related research and open-source links associated with the datasets and toolkits discussed in this section.

### 6.2 Evaluation Metrics

As stated in Section 3.2, the objective of failure diagnosis in current microservice systems is to localize the root cause and classify the type of failure. Root cause localization [16, 17, 19–22, 26, 28–32, 32, 33, 33–36, 38, 43, 45–48, 50, 52, 55–58, 64, 67, 69, 70, 74–76, 90–92, 95, 96, 105, 105, 108, 120] is typically evaluated with *Accuracy@k* (*A@k*), *average Accuracy@K* (*Avg@K*), *Precision@K* (*PR@K*), and *mean average PR@K* (*MAP@K*), which are using the following commonly used ranking evaluation metrics in recommendation systems. Given $A$ as the set of system failures, $a$ as one failure in $A$, $V_a$ as the real root cause of $a$, and $R_a[k]$ as the predicted top-k set of $a$. These metrics are defined as $A@k = \frac{1}{|A|} \sum_{a \in A} \begin{cases} 1, & \text{if } V_a \in R_a[k] \\ 0, & \text{otherwise} \end{cases}$, $PR@k = \frac{1}{|A|} \sum_{a \in A} \frac{\sum_{i<k} R_a[i] \in V_a}{min(k,|V_a|)}$, $Avg@K = \frac{\sum_{1 \le k \le K} A@k}{K}$ and $MAP@K = \frac{1}{K|A|} \sum_{a \in A} \sum_{1 \le k \le K} PR@k$. They indicate the probability of the real root cause being present in the predicted top-k set. In practice, operators usually examine the top-5 results. In addition, *mean reciprocal rank* (*MRR*) and *ranking percentile* (*RP*) are defined as $MRR = \frac{1}{A} \sum_{a \in A} \frac{1}{rank(V_a, R_a)}$ and $RP = (1 - \frac{rank(V_a, R_a)}{num(R_a)}) \times 100\%$, where $rank(V_a, R_a)$ is the ranking position of $V_a$ in $R_a$, $num(R_a)$ is the number of suspicious root causes in $R_a$. They are used to measure the ranking capability of the models. Larger values of these metrics indicate higher-ranking predictions and better diagnostic performance of the models.

Table 9. A list of publicly available datasets and toolkits.

| Name | | Related Research | Link |
|---|---|---|---|
| **Datasets** | AIOps Challenge 2020 | [21, 55, 58, 73, 74, 76] | https://github.com/NetManAIOps/AIOps-Challenge-2020-Data |
| | AIOps Challenge 2021 | [73, 147, 148] | https://www.aiops.cn/gitlab/aiops-nankai/data/trace/aiops2021 |
| | AIOps Challenge 2022 | [57, 147, 148] | https://competition.aiops-challenge.com/home/competition |
| | GAIA | [19, 138, 145] | https://github.com/CloudWise-OpenSource/GAIA-DataSet |
| | TrainTicket-DéjàVu | [92] | https://zenodo.org/records/6955909 |
| | TrainTicket-Eadro | [51] | https://github.com/BEbillionaireUSD/Eadro |
| | TrainTicket-Nezha | [96] | https://github.com/IntelligentDDS/Nezha/tree/main/rca_data |
| | Loghub | - | https://github.com/logpai/loghub |
| | SWaT, WADI | [32, 33] | https://itrust.sutd.edu.sg/itrust-labs_datasets/dataset_info/ |
| | SMAP, MSL | - | https://github.com/khundman/telemanom |
| **Toolkits** | DISTALYZER | [86, 154] | http://www.macesystems.org/distalyzer,Lu2017LogbasedAT |
| | FDiagV3 | [87, 155, 156, 156−158] | http://diag-toolkits.github.io/FDiag |
| | Log3C | [60, 83, 159] | https://github.com/logpai/Log3C |
| | SwissLog | [60, 160] | https://github.com/IntelligentDDS/SwissLog |
| | LogKG | [107, 161] | https://anonymous.4open.science/r/LogKG-A6BD |
| | DBSherlock | [94, 97, 98, 162] | http://dbseer.org |
| | Sieve | [21, 29, 34, 40, 45−48, 52, 54−56, 69, 70, 90] | https://sieve-microservices.github.io/ |
| | DyCause | [48, 50, 51, 67, 70, 163] | https://github.com/PanYicheng/dycause_rca |
| | DéjàVu | [19, 51, 92, 138, 145, 147, 148, 164] | https://github.com/NetManAIOps/DejaVu |
| | CIRCA | [32, 42, 51, 96] | https://github.com/NetManAIOps/CIRCA |
| | RootCLAM | [44] | https://github.com/hanxiao0607/RootCLAM |
| | TraceAnomaly | [21, 22, 26, 52, 55, 56] | https://github.com/NetManAIOps/TraceAnomaly.git |
| | TraceRCA | [48, 56, 57, 67, 96, 147] | https://github.com/NetManAIOps/TraceRCA.git |
| | MicroCBR | [16, 138, 145, 148] | https://github.com/Fengrui-Liu/MicroCBR |
| | UniDiag | [138] | https://github.com/AIOps-Lab-NKU/UniDiag |
| | Eadro | [19, 51, 148, 161, 165, 166] | https://github.com/BEbillionaireUSD/Eadro |
| | Medicine | [145] | https://github.com/AIOps-Lab-NKU/Medicine |
| | DiagFusion | [19, 96, 138, 145, 147, 148, 161, 164, 165] | https://github.com/AIOps-Lab-NKU/DiagFusion |
| | Nezha | [96, 164, 167] | https://github.com/IntelligentDDS/Nezha |
| | DeepHunt | [147] | https://github.com/bbyldebb/DeepHunt |
| | ART | [148] | https://github.com/bbyldebb/ART |

Apart from the aforementioned commonly used metrics, *exam score* (*ES*) refers to the average number of incorrect candidate options that operators must manually exclude before diagnosing the real root cause of each failure.

Failure classification mostly utilizes evaluation metrics of multi-class tasks in machine learning to demonstrate diagnostic effectiveness. Multi-class tasks can independently evaluate each class using binary classification methods to obtain multiple binary evaluation metrics. These metrics, including *precision* (*P*), *recall* (*R*), and $F_1 - score$ ($F_1$), can accurately reflect the results of each class. With *true positives* (*TP*), *false positives* (*FP*), and *false negatives* (*FN*), the calculation for binary classification tasks is given by $P = \frac{TP}{TP+FP}$, $R = \frac{TP}{TP+FN}$, $F_1 = \frac{2 \times P \times R}{(P+R)}$. Based on the above, the metrics for failure classification can be obtained [17, 19, 39, 44, 59, 68, 83, 84, 98, 98, 106, 168, 169]. *Micro average* is calculated by summing up the $TP$, $FP$, and $FN$ across all classes, where *Micro Precision = Micro Recall = Micro $F_1$ − score* $= \frac{\sum_{i=1}^{n} TP_i}{\sum_{i=1}^{n} (TP_i+FP_i)} = \frac{\sum_{i=1}^{n} TP_i}{\sum_{i=1}^{n} (TP_i+FN_i)}$. As *Micro average* cannot differentiate between different classes, it is more suitable for scenarios with uneven sample distribution.

*Macro average* is calculated by computing $P$, $R$, and $F_1$ for each class, and taking the average, where *Macro Precision* $= \frac{1}{n} \sum_{i=1}^{n} P_i$, *Macro Recall* $= \frac{1}{n} \sum_{i=1}^{n} R_i$ and *Macro $F_1$ − score* $= \frac{1}{n} \sum_{i=1}^{n} F_{1i}$. As *Macro average* assigns the same weight to each class, it can treat each class equally and thus can be influenced by rare classes.

When the samples are imbalanced, it is not appropriate to assign the same weight to each class. *Weighted average* is calculated by computing $P$, $R$, and $F_1$ for each class, and then assigning different weights based on the sample size of each class, where *Weighted Precision* $= \sum_{i=1}^{n} P_i \times W_i$, *Weighted Recall* $= \sum_{i=1}^{n} R_i \times W_i$ and *Weighted $F_1$ − score* $= \sum_{i=1}^{n} F_{1i} \times W_i$.

Some failure diagnosis solutions [98] that employ clustering methods also utilize *clustering accuracy* ($AC$) and *normalized mutual information* ($NMI$) , which are good measures of clustering quality.

# 7 RELATED WORK

Failure diagnosis is crucial in microservice systems, and thanks to the continuous attention of researchers, significant progress has been made in this domain. Relevant surveys on this topic have been published successively. However, previous studies mainly focus on a single task and don't comprehensively update and review existing techniques.

Oliner et al. [10] survey focus on log analysis in computer-system logs. Log analysis can help optimize or debug system performance. Wong et al. [5] outline techniques for localizing the root causes of failures in the source code of an individual software program. These techniques are classified into eight categories, including slice-based, spectrum-based, statistics-based, program state-based, machine learning-based, data mining-based, model-based, and miscellaneous. Before 2016, microservices architecture had not received widespread attention, and most applications could not escape the constraints of traditional monolithic software, only being able to refactor application code based on the original traditional software. Therefore, the studies surveyed by Oliner et al. [10] and Wong et al. [5] are still based on the failure diagnosis techniques of the traditional software architecture, with limitations in improving and applying them. Gao et al. [6, 7] classify failure diagnosis techniques into core methods, including model-based, signal-based, knowledge-based, and hybrid/active diagnosis techniques. Sole et al. [8] survey available root cause analysis models and the corresponding generation and inference algorithms in distributed systems from multiple performance evaluation perspectives (*i.e.*, scalability and real-time reaction). However, the studies surveyed by Gao et al. [6, 7] and Sole et al. [8] overlook specific modalities and granularity. Additionally, the studies [5–7, 10] are dated up to 2016 and do not reflect the current advances in failure diagnosis techniques.

He et al. [11] provide a detailed overview of automated log analysis in large-scale software systems, including how to employ logs to detect anomalies, predict failures, and facilitate diagnosis. He et al. [11] also survey open-source datasets and toolkits. Li et al. [13] present an industrial survey of microservice tracing and analysis, which surveyed different root cause analysis techniques including visualization, statistics (*i.e.*, statistical calculation of related metrics), and rule-based decision. However, Li et al. [13] instead survey more advanced root cause analysis techniques, including machine learning and data mining. Furthermore, the studies surveyed by He et al. [11] and Li et al. [13] only consider logs or traces and do not qualitatively compare the different techniques, while we not only classify them based on the overall method in detail but also qualitatively analyze and discuss them.

The above studies either overlook specific modalities [6–8] or only focus on one modality, such as logs [5, 10, 11] or traces [13]. The studies surveyed by Notaro et al. [12] and Soldani et al. [9] are closer to ours, considering single-modal data (i.e., using either metrics, logs, or traces). Notaro et al. [12] focus on failure management of the applicability requirements and quantitative results, including failure diagnosis. Notaro et al. [12] consider two tasks of failure diagnosis and provide explanations and discussions for each. However, Notaro et al. [12] do not classify the core methods in fine granularity, which could not help operators understand the processes and paradigms of existing failure diagnosis techniques. Soldani et al. [9] provide a structured overview and qualitative analysis of existing anomaly detection and root cause analysis in large-scale software systems. Soldani et al. [9] mainly survey direct and graph-based root cause analysis techniques, while overlooking some feature-based techniques [16, 17, 19, 30, 39, 43, 44, 51–53, 57–59, 62, 63, 73, 74, 80, 83, 86–90, 92, 93, 96, 98, 101, 103, 104, 107, 120, 129]. In addition, we not only analyze these techniques from the perspectives of granularity, explainability, accuracy, and costs but also complement qualitative evaluations of characteristics and portability. We also focus on publicly available datasets and toolkits, summarizing the evaluation metrics for failure diagnosis tasks, aiming to support operators in deploying and validating these techniques.

There are also some noteworthy empirical studies and evaluations by Zhou et al. [102], Arya et al. [170] and Garg et al. [171]. Zhou et al. [102] conduct an industrial survey focusing on typical failure analysis in microservice systems. The experimental results demonstrate that using proper tracing and visualization techniques can help in diagnosing various failures related to microservice interactions. Arya et al. [170], using the aforementioned TrainTicket microservice system, evaluate the performance of various state-of-the-art Granger causal inference techniques. Garg et al. [171] comprehensively evaluated algorithms for anomaly detection and diagnosis in modern cyber-physical systems through training 11 deep learning models on 7 multivariate time series datasets. Therefore, the studies [102, 170, 171] complement our qualitative comparison, quantitatively comparing the performance and accuracy of existing techniques. However, these studies [102, 170, 171] also lack comparisons of failure diagnosis techniques based on multimodal data.

## 8 CONCLUSIONS

As microservice systems grow in scale and complexity, and deployment techniques advance, failure diagnosis has emerged as a critical frontier. Our survey is the first to comprehensively examine failure diagnosis techniques using multimodal data (i.e., logs, metrics, traces, events and topology) in microservice systems, covering both single-modal and multimodal fusion approaches from 2003 to present. We summarize the characteristics, trends, and research progress of existing solutions for researchers and practitioners. Our qualitative analysis examines various aspects including granularity, explainability, characteristics, portability, accuracy, and costs, aiming to promote stable microservice system development. While surveyed papers often measure diagnostic accuracy and compare solutions, we recognize that other factors like training data scale, efficiency, storage, and invocation overhead are also crucial. We recommend combining qualitative and quantitative analyses to guide practitioners in selecting appropriate solutions for real-world applications. Additionally, we offer the first survey to address problem statement, evaluation metrics, publicly available datasets, and toolkits specific to failure diagnosis. We aim to contribute to the development of modern microservice systems, provide researchers with comprehensive background and references, and inspire future studies in failure diagnosis techniques.

## ACKNOWLEDGMENTS

This work is supported by the Advanced Research Project of China (No. 31511010501), and the National Natural Science Foundation of China (62272249, 62302244).